\documentclass[superscriptaddress,secnumarabic,twocolumn,
 amssymb,amsmath,nobibnotes,aps,prd,showkeys,showpacs,nofootinbib]{revtex4}
\usepackage{amsmath}
\usepackage{amssymb}
\usepackage{mathtools}
\usepackage{graphicx}
\usepackage{color}
\usepackage[normalem]{ulem}
\usepackage[utf8]{inputenc}

\newcommand{\N}{\mathbb{N}}

\newcommand{\abs}[1]{\left|{#1}\right|}

\DeclareMathOperator{\defeq}{\vcentcolon =}

\newcommand{\der}[2]{\frac{\mathrm{d}{#1}}{\mathrm{d}{#2}}}
\newcommand{\pder}[3][]{\frac{\partial^{#1}{#2}}{\partial{#3}^{#1}}}

\renewcommand{\d}[1]{\mathrm{d}{#1}}
\DeclareMathOperator{\denom}{\mathcal{D}}

\newcommand{\HHT}{\Upsilon}

\newif\ifimages
\imagestrue

\begin{document}

\title{Growth of resonances and chaos for a spinning test particle in the Schwarzschild background}
\author{Ond\v{r}ej Zelenka}
\email{ondrej.zelenka@uni-jena.de}
\affiliation{Astronomical Institute of the Academy of Sciences of the Czech Republic,
Bo\v{c}n\'{i} II 1401/1a, CZ-141 00 Prague, Czech Republic}
\affiliation{Theoretisch-Physikalisches Institut,  Friedrich-Schiller-Universit{\"a}t Jena, 07743, Jena, Germany}
\author{Georgios Lukes-Gerakopoulos}
\email{gglukes@gmail.com}
\affiliation{Astronomical Institute of the Academy of Sciences of the Czech Republic,
Bo\v{c}n\'{i} II 1401/1a, CZ-141 00 Prague, Czech Republic}
\author{Vojt\v{e}ch Witzany}
\email{witzany@asu.cas.cz}
\affiliation{Astronomical Institute of the Academy of Sciences of the Czech Republic,
Bo\v{c}n\'{i} II 1401/1a, CZ-141 00 Prague, Czech Republic}
\author{Ond\v{r}ej Kop\'{a}\v{c}ek}
\email{kopacek@ig.cas.cz}
\affiliation{Astronomical Institute of the Academy of Sciences of the Czech Republic,
Bo\v{c}n\'{i} II 1401/1a, CZ-141 00 Prague, Czech Republic}

\begin{abstract}
Inspirals of stellar mass compact objects into supermassive black holes are known as extreme mass ratio inspirals. In the simplest approximation, the motion of the compact object is modeled as a geodesic in the space-time of the massive black hole with the orbit decaying due to radiated energy and angular momentum, thus yielding a highly regular inspiral. However, once the spin of the secondary compact body is taken into account, integrability is broken and prolonged resonances along with chaotic motion appear. 

We numerically integrate the motion of a spinning test body in the field of a non-spinning black hole and analyse it using various methods. We show for the first time that resonances and chaos can be found even for astrophysical values of spin. On the other hand, we devise a method to analyse the growth of the resonances, and we conclude that the resonances we observe are only caused by terms quadratic in spin and will generally stay very small in the small-mass-ratio limit. Last but not least, we compute gravitational waveforms by solving numerically the Teukolsky equations in the time-domain and establish that they carry information on the motion's dynamics. In particular, we show that the time series of the gravitational wave strain can be used to discern regular from chaotic motion of the source.
\end{abstract}

\pacs{~~}
\keywords{Gravitation, Gravitational waves; Dynamical systems}
\maketitle

\section{Introduction}

The last few years have witnessed the birth of gravitational wave astronomy. Although the first indirect evidence of gravitational waves has been known since the early 1980s \cite{Weisberg81}, the first direct detection GW150914 came by the LIGO Scientific Collaboration \cite{Abbott16}.  Since then, $\sim 10$ more events have been confirmed \cite{LIGO18}. All of them have been calculated to be mergers of compact binaries of comparable masses between $10^1 M_\odot$ and $10^2 M_\odot$.

The Laser Interferometric Space Antenna (LISA) mission, the first dedicated space-based gravitational wave detector, is expected to be launched in 2034 \cite{Babak17}. Unlike ground-based detectors, the sensitivity of LISA will lie in the range from $10^{-4}\:\mathrm{Hz}$ to $10^{-2}\:\mathrm{Hz}$, allowing for observation of many different types of sources such as supermassive black hole mergers \cite{Klein15} or stellar mass compact binaries long before merger \cite{Sesana16}. Another exciting prospect is to use these sources as independent standard sirens that allow us to constrain cosmological models \cite{Tamanini16}. One of the prime targets are also the inspirals of stellar mass ($10^{0}\: M_\odot - 10^{1}\: M_\odot$) compact objects (black holes or neutron stars) into supermassive black holes ($10^{5}\: M_\odot - 10^{10}\: M_\odot$), which are called \textit{extreme mass ratio inspirals} (EMRIs). It is expected that during its planned 4-10 year mission, LISA will detect between 1 and 2000 EMRIs per year \cite{Babak17}.

Predicting the evolution of a compact binary requires solving the full nonlinear set of Einstein equations. Since this is essentially impossible to do exactly, an obvious choice is to use numerical relativity simulations instead, which, on the other hand, come with a hefty computational price tag. Even though the numerical computation of gravitational-wave inspirals and mergers have been achieved in the case of comparable-mass binaries \cite{duez2018}, in the case of an EMRI the corresponding computational cost is prohibitive due to the huge separation of temporal and spatial scales involved in the problem.

Instead, in the extreme mass ratio limit $q = \mu/M \ll 1$, where $\mu$ and $M$ are the respective masses of the secondary and primary compact object, one can model the dynamics of the inspiral as the motion of a ``particle'' on the background of the primary black hole and employ black hole perturbation theory to compute the waveform and radiation-reaction forces on the particle itself \cite{Barack2018}. The force on the test particle is then not only the gravitational self-force due to the perturbation caused to the background, but also finite-size effects; forces that come about due to the fact that the real object in this model is not truly a point particle.

In this work we focus on the finite size effects at the dipole level, which amounts to the model known as a ``spinning particle''. From the multipole expansion of the secondary object \cite{Dixon70a,Dixon70b,Dixon74} we ignore the quadrupole and all the higher multipole moments. Moreover, we treat the secondary as a test object, so we ignore also the dynamical aspects of the self-force. To reproduce the gravitational radiation emitted by the secondary object, we solve numerically the Teukolsky equations \cite{Teukolsky73} in the time domain.

To model this secondary spinning test object, we use the Mathisson-Papapetrou-Dixon (MPD) equations \cite{Mathisson37,Papapetrou51,Dixon70a,Dixon70b,Dixon74}. These equations correspond in general to a non-integrable system, which exhibits chaotic behavior and prolonged resonances \cite{Suzuki96,Hartl02}. However, when MPD are linearized in spin, it appears that under certain conditions they are approximately separable  \cite{Witzany19}. It is unclear whether this means that chaos and prolonged resonances do not appear at linear order in spin \cite{Kusnt16,Ruangsri16}. Whether spin-induced chaos and resonances play any role in EMRI needs yet to be determined in order to make accurate predictions for LISA waveforms. 

In this work, we consider the \textit{non-linearized} in spin MPDs to describe the motion of a spinning particle in a curved spacetime; we restrict ourselves to the motion of an object moving in a non-spinning supermassive black hole background, described by the Schwarzschild spacetime. We use Poincar\'{e} sections, the rotation number and recurrence analysis to study the motion of the particle and to demonstrate the emergence of chaos. Finally, to establish a link between the dynamical features of the spinning particle motion and the resulting gravitational waveform, we employ the recurrence analysis on the waveforms.

The organization of the paper is as follows: section~\ref{sec:mpd} summarizes some basic theoretical results concerning the equations of motion of spinning particles, regular and chaotic behavior and the Schwarzschild spacetime. Section~\ref{sec:aa} demonstrates how resonances grow with spin. In section~\ref{sec:waves}, we compute gravitational waveforms from the studied system and in section~\ref{sec:rps}, we use recurrence analysis to establish a link between chaos in the motion and in the waveforms. Section~\ref{sec:concl} summarizes the work.

We will use Greek indices $\mu, \nu, ... = t, r, \theta, \phi$ to denote coordinate components of tensors and capital latin indices $A, B, ... = 0, 1, 2, 3$ to denote tetrad components. The metric signature convention used is $\left(-,+,+,+\right)$. A dot above a symbol denotes the absolute derivative with respect to the proper time $\dot{A}^{\mu\hdots}_{\nu\hdots} = \mathrm{D} A^{\mu\hdots}_{\nu\hdots}/ \d\tau$ and indices separated by a comma or semicolon denote a partial or covariant derivative, respectively. The sign convention for the Riemann tensor is that  
$2\omega_{\rho;[\mu\nu]}  \defeq \omega_\sigma {R^\sigma}_{\rho\mu\nu}$ for any covector field $\omega$.

\section{Spinning particles in the Schwarzschild spacetime}\label{sec:mpd}

\subsection{Equations of motion for spinning particles}

We model the EMRI on time-scales shorter than the radiation-reaction timescale as the motion of an extended body in the fixed spacetime background of a static black hole. This motion can be described by a comprehensive multipole formalism developed by Dixon \citep{Dixon70a, Dixon70b, Dixon74}. If one concerns oneself with only the gravitational interaction and restricts the multipole expansion of the extended body to the monopole and dipole terms by neglecting the quadrupole and higher-order moments, the MPD equations of motion reduce to 
\begin{subequations}\label{eq:mpd}
\begin{align}
    \dot{P}^\mu &= -\frac{1}{2}{R^\mu}_{\nu\rho\sigma}u^\nu S^{\rho\sigma} \:,\\
    \dot{S}^{\mu\nu} &= 2P^{[\mu}u^{\nu]} = P^\mu u^\nu - P^\nu u^\mu \: \label{eq:mpdspin},
\end{align}
\end{subequations}
where $S^{\mu\nu}$ is the spin-tensor, $P^\mu$ is the four-momentum and $u^\mu = \d x^\mu/ \d\tau$ is the four-velocity.  The worldline of a particle $x^\mu (\tau)$ is parameterized by the proper time defined by
\begin{equation}
    \d\tau^2 \defeq -g_{\mu\nu}\d x^\mu\d x^\nu \:.
\end{equation}
Neglecting the dipole term $(S^{\mu\nu})$ in the expansion would yield geodesic motion.

If we define the mass with respect to the four-velocity
\begin{equation}\label{eq:massU}
    m \defeq -P_\mu u^\mu
\end{equation}
and the mass with respect to the four-momentum
\begin{equation}\label{eq:massP}
    \mu \defeq \sqrt{-P_\mu P^\mu} \:,
\end{equation}
then Eq.~\eqref{eq:mpdspin} can be rewritten as
\begin{equation}
    P^\mu = m u^\mu - u_\sigma \dot{S}^{\mu\sigma} \:.
\end{equation}
Another useful quantity is the spin magnitude
\begin{equation}\label{eq:spinm}
    S \defeq \sqrt{\frac{1}{2}S_{\mu\nu}S^{\mu\nu}} \:.
\end{equation}
For a continuous symmetry of the spacetime background with the corresponding Killing vector field $\xi$, this system admits an integral of motion in the form \cite{Semerak99}
\begin{equation}\label{eq:int_mpd}
    C(\xi) = P_{\sigma}\xi^{\sigma} - \frac{1}{2}\xi_{\rho;\sigma}S^{\rho\sigma} \:.
\end{equation}

\subsubsection{Spin supplementary condition}

The MPD equations do not fully determine the system's evolution. Therefore, it is necessary to specify an additional spin supplementary condition (SSC)
\begin{equation}\label{eq:ssc_gen}
    V_\mu S^{\mu\nu} \defeq 0 \:,
\end{equation}
where $V$ is a time-like vector. The meaning of a SSC is related to the multipole formalism \cite{Dixon70a}: one describes the extended body using its multipole moments and $V$ in Eq. \eqref{eq:ssc_gen} determines the frame used to calculate these moments. Several different SSCs have been proposed (see, e.g., \cite{Witzany18, LukesGerakopoulos17a}); however, the existence of a four-vector $V$ such that Eq. \eqref{eq:ssc_gen} imposes already the constraint  $\epsilon_{\mu\nu\rho\sigma}S^{\mu\nu}S^{\rho\sigma} = 0$ on the system.

We choose the Tulczyjew-Dixon (TD) SSC \cite{Tulczyjew65,Dixon70a}
\begin{equation}\label{eq:td_ssc}
    P_\mu S^{\mu\nu} = 0 \:.
\end{equation}
Due to this condition, the mass~\eqref{eq:massP} and the spin magnitude~\eqref{eq:spinm} are integrals of motion \cite{Semerak99}. The system as given above together with the SSC leads to the equation \cite{Ehlers77}
\begin{equation}
    u^\mu = \der{x^\mu}{\tau} = \frac{m}{\mu^2}\left(P^\mu + \frac{2S^{\mu\nu}R_{\nu\gamma\kappa\lambda}P^\gamma S^{\kappa\lambda}}{4\mu^2+R_{\chi\eta\omega\xi}S^{\chi\eta}S^{\omega\xi}}\right) \:.
\end{equation}
Since $u^\mu u_\mu=-1$ the mass~\eqref{eq:massP} can be expressed as
\begin{equation}
    m = \frac{\mathcal{A}\mu^2}{\sqrt{\mathcal{A}^2\mu^2 - \mathcal{B}S^2}} \:,
\end{equation}
where
\begin{subequations}
\begin{align}
    \mathcal{A} &= 4\mu^2 + R_{\alpha\beta\gamma\delta}S^{\alpha\beta}S^{\gamma\delta} \:, \\
    \mathcal{B} &= 4h^{\kappa\eta}R_{\kappa\iota\lambda\mu}P^\iota S^{\lambda\mu}R_{\eta\nu\omega\pi}P^\nu S^{\omega\pi} \:, \\
    {h^\kappa}_\eta &= \frac{1}{S^2}S^{\kappa\rho}S_{\eta\rho} \:.
\end{align}
\end{subequations}
Thus, the derivatives $\dot{x}^\mu$, $\dot{P}^\mu$ and $\dot{S}^{\mu\nu}$ are uniquely expressed and the system of ordinary differential equations (ODEs) is complete.

\subsubsection{Canonical formalism}

While the system as discussed so far is a well-defined dynamical system, it is simply a set of ODEs, whereas the symplectic structure of a Hamiltonian system would provide us with tools to identify reducible degrees of freedom (DoF) and eliminate them from the ODE system. By reducing the degrees of freedom we can easier understand and study dynamical features of the system. We will use the formalism described in \cite{Witzany18}. We only provide a very concise summary here; for more details see \cite{Witzany18}.

We take an orthonormal tetrad $e^{A\mu}$, $g_{\mu\nu}e^{A\mu} e^{B\nu} = \eta^{AB}$ and define the "dual" tetrad as ${e^A}_\mu \defeq g_{\mu\nu}e^{A\nu}$. The new phase space coordinates are defined by
\begin{subequations}
\begin{align}
    x^\mu & \:, \\
    p_\mu &\defeq P_\mu + \frac{1}{2}{e}_{A\nu;\mu}{e_B}^{\nu} S^{AB} \:, \\
    S^{AB} &\defeq {e^A}_\mu {e^B}_\nu S^{\mu\nu} \:.
\end{align}
\end{subequations}
With the Poisson brackets
\begin{subequations}\label{eq:poisson}
\begin{align}
    \{x^\mu,p_\nu\} &= \delta^\mu_\nu \:, \\
    \begin{split}
    \{S^{AB},S^{CD}\} &=  \eta^{AC}S^{BD} - \eta^{AD}S^{BC} \\
                      &+ \eta^{BD}S^{AC} - \eta^{BC}S^{AD} \:,
    \end{split} \\
    \begin{split}
    \{x^\mu,x^\nu\} &= \{p_\mu,p_\nu\} = \\
    = \{x^\mu,S^{AB}\} &= \{p_\mu,S^{AB}\} = 0 \:,
    \end{split}
\end{align}
\end{subequations}
the Hamiltonian
\begin{equation}\label{eq:ham_mpd}
    H = \frac{m}{2\mu^2}\left[P_\mu P^\mu - \frac{4S^{\nu\gamma}{R^\mu}_{\gamma\kappa\lambda}S^{\kappa\lambda}P_\mu P_\nu}{4\mu^2+R_{\chi\eta\omega\xi}S^{\chi\eta}S^{\omega\xi}} + \mu^2\right] \:.
\end{equation}
generates the system of MPD equations with the TD SSC. By a further transformation, one obtains canonical variables \cite{Witzany18}; for the purposes of this work, the existence of these canonical set of variables is enough. One of the interesting realizations of this is that the addition of the spin to the geodesic system in the case of the TD SSC leads to only one additional DoF \cite{Witzany18}. 

\subsection{Dynamical systems}

This section summarizes elements of non-linear dynamical systems. For a more comprehensive insight into the topic, see \cite{Lichtenberg92, Meiss92, Morbidelli02}.

\subsubsection{Integrable systems}\label{sec:integrability}

An important notion is that of an integrable system, which is closely connected to integrals of motion. Let us consider an autonomous Hamiltonian system with $N$ DoF, i.e. a $2N$-dimensional phase space, and Hamiltonian $H$. Let us also assume that there exist $n$ non-trivial linearly independent integrals of motion $I_i,\: i=1,2,\hdots,n$ in involution ($\{I_i, I_j\}=0$). Then, a canonical transformation exists which equates several momenta with the integrals and the Hamiltonian is thus independent of the corresponding positions. Then the evolution of the remaining phase space variables is independent of the separated DoF and we call it the \textit{reduced system}. 

If there are $N$ independent integrals of motion in involution, then according to Liouville theorem, there exists a set of variables $\theta^i$, $I_i$ such that:
\begin{subequations}\label{eq:lin_variables}
\begin{align}
    \theta^i\left(\tau\right) &= \theta^i\left(0\right) + \tau\omega^i\left(I_j\right) \:,\\
    I_i\left(\tau\right) &= I_i\left(0\right) \:,
\end{align}
\end{subequations}
and the system is called \textit{integrable}. In addition, if the motion is bounded, then it lies on a nested family of $N$-dimensional tori (see, e.g., \cite{Lichtenberg92}). The $\theta^i$ variables are then typically $2\pi$-periodic and the $\omega^i$'s are the characteristic frequencies.

The relevant system in this work possesses enough integrals to be reduced to two DoF, so from now on, we will restrict ourselves to systems with $N=2$ and bounded motion. In an integrable system, the ratio of characteristic frequencies is known as \textit{rotation number} $\omega = \omega^1/\omega^2$. When $\omega \in \mathbb{Q}$, the motion is periodic, the torus is called \textit{resonant} and the corresponding phase space region is called a \textit{resonance}. On the other hand, if $\omega\in\mathbb{R}\setminus\mathbb{Q}$, the orbit densely covers the whole torus (this is called quasiperiodic motion).

A very powerful phase space visualization tool is the \textit{Poincar\'{e} section}. Orbits lie on a hypersurface of constant Hamiltonian in the phase space. In it, one can choose a 2-dimensional surface, which is transversal to the Hamiltonian flow. We call such a surface the \textit{surface of section} or Poincar\'{e} section. By considering succesive intersections of a trajectory with the surface of section, the original 2-DoF continuous-time system is converted to a 1-DoF discrete-time system. In the case of an integrable system, all intersections corresponding to a single non-resonant trajectory will lie on a closed curve (usually called an invariant circle). For a resonant torus with $\omega=r/s$, a single trajectory will only form a finite set of $s$ periodic points in the surface of section. 

A Poincar\'{e} section provides a practical method to evaluate the rotation number. One must identify the period-1 point $\vec{x}_c$ on the surface of section and take angles between successive intersections $\vec{x}_i$ with the surface of section with respect to $\vec{x}_c$ as
\begin{equation}
    \vartheta_i \defeq \mathrm{ang}\left[\left(\vec{x}_{i+1}-\vec{x}_c\right), \left(\vec{x}_{i}-\vec{x}_c\right)\right] \:.
\end{equation}
Then the rotation number, up to an additive integer, can be determined as
\begin{equation}\label{eq:angular_moment}
    \nu_\vartheta \defeq \lim_{n\to\infty} \frac{1}{2\pi n}\sum_{i=1}^n \vartheta_i \:.
\end{equation}
If the trajectory lies on an invariant torus, then $\nu_\vartheta = \omega \mod 1$ \cite{Voglis98}. The error of the limit for finite $n$ is bounded by $1/n$. The rotation number for typically changes monotonically for initial conditions along a direction getting further away from $\vec{x}_c$.

\subsubsection{Non-integrable systems}\label{sec:perturbation}

The structure as described in the section~\ref{sec:integrability} fully applies only in the integrable case. Upon application of a small perturbation, one typically gets a non-integrable system, which, however, retains a lot of structure from the integrable system. The surviving invariant circles are those that are "sufficiently far enough from resonances" according to the KAM theorem (see, e.g., \cite{Meiss92,Lichtenberg92}). The set bounded by the outermost surviving invariant circle is then called the \textit{main island of stability}.

Regarding the resonances after the perturbation, the Poincar\'{e}-Birkhoff theorem (see, e.g., \cite{Lichtenberg92}) states that from a resonance $\omega = r/s$ in the integrable system, $2ns$ of the periodic points survive in the perturbed system, where $n \in \N$. Half of the surviving points are stable (elliptic), around which \textit{islands of stability} arise, and the other half unstable (hyperbolic). The dynamics near an unstable point $\vec{x}_f$ is related to invariant manifolds, wherein the stable and unstable manifold contain points which tend to $\vec{x}_f$ in forward time and reversed time, respectively. Same type of manifolds cannot intersect themselves, e.g. a stable manifold cannot intersect another stable, but stable manifolds can intersect unstable ones. On these manifolds and their intersections move the \textit{chaotic} orbits, which fill densely a two-dimensional subset of the surface of section.

The general picture of the perturbed phase space on a Poincar\'{e} section is as follows: there remains a central fixed point with many quasi-periodic KAM circles, forming the main island of stability. In the resonances arise chaotic regions, densely filled by a single orbit, and islands of stability formed around the stable periodic points of the resonances. These are remarkably similar to the main island of stability, leading to tertiary islands of stability etc. At every order, one can find higher-order resonances, elliptic points with islands of stability and hyperbolic points with homoclinic (consecutive sections between manifolds corresponding the same periodic point) and heteroclinic (consecutive sections between manifolds corresponding the different periodic points)  orbits. 

The quantity $\nu_\vartheta$ defined in Eq. \eqref{eq:angular_moment} can also be helpful to visualize the phase space of a non-integrable system. If one chooses a parametrized line of initial conditions in the surface of section, computes the $\nu_\vartheta$ for each of them and plots it as a function of the initial conditions, one obtains the \textit{rotation curve}. If the initial conditions are chosen so that they only cross invariant circles in one direction, then the rotation curve is monotonous. When passing through a resonance, $\nu_\vartheta$ forms a plateau in the rotation curve. Multiple initial conditions in a chaotic region, however, lead to unpredictable values and wildly differ. Thus, one can detect resonances in a non-integrable system by either a plateau or non-monotonous variations in the rotation curve.

\subsection{In the Schwarzschild spacetime}

\subsubsection{Schwarzschild spacetime}
The Schwarzschild metric, describing a non-spinning black hole of mass $M$, is in Schwarzschild coordinates given by the line element
\begin{equation}\label{eq:schw}
    \begin{split}
    &g_{\mu\nu}\d x^\mu\d x^\nu = \\
    & -f\left(r\right)\d t^2 + \frac{1}{f\left(r\right)}\d r^2 + r^2\left(\d \theta^2 + \sin^2\!\theta\d \phi^2\right) \:,
    \end{split}
\end{equation}
where
\begin{equation}
    f\left(r\right) = 1 - \frac{2M}{r} \:.
\end{equation}
The spacetime is stationary and spherically symmetric, i.e. there exist a timelike Killing vector field and three spacelike Killing vector fields 
\begin{subequations}\label{eq:killing}
\begin{align}
    \xi_{(t)} &= \pder{}{t} \:, \\
    \xi_{(x)} &= -\sin\phi\pder{}{\theta} - \cos\phi\cot\theta\pder{}{\phi} \:, \\
    \xi_{(y)} &= \cos\phi\pder{}{\theta} - \sin\phi\cot\theta\pder{}{\phi} \:, \\
    \xi_{(z)} &= \pder{}{\phi} \:.
\end{align}
\end{subequations}

\subsubsection{Integrals of motion}

The formula for integrals of motion in Eq. \eqref{eq:int_mpd} can be expressed for computational convenience using either the $P_\mu$ or the canonical $p_\mu$ as
\begin{equation}\label{eq:int_simple}
\begin{split}
    C(\xi) = P_{\sigma}\xi^{\sigma} - \frac{1}{2}g_{\rho\alpha}{\xi^\alpha}_{,\sigma}S^{\rho\sigma} - \frac{1}{2}g_{\beta\rho,\sigma}\xi^\beta S^{\rho\sigma} = \\ p_{\sigma}\xi^{\sigma} - \frac{1}{2}g_{\rho\alpha}{\xi^\alpha}_{,\sigma}S^{\rho\sigma} - \frac{1}{2}g_{\alpha\nu}\xi^\mu {{e_A}^\alpha}_{,\mu} {e_B}^\nu S^{AB} \:.
\end{split}
\end{equation}
Together with the tetrad
\begin{subequations}
\begin{alignat}{3}
    e_0 &= \frac{1}{\sqrt{f}}\pder{}{t} \:, \quad && e_1 &&= \sqrt{f}\pder{}{r} \:, \\
    e_2 &= \frac{1}{r}\pder{}{\theta} \:, \quad && e_3 &&= \frac{1}{r\sin\theta}\pder{}{\phi} \:,
\end{alignat}
\end{subequations}
the Killing fields in Eqs. \eqref{eq:killing} give rise to the integrals
\begin{subequations}\label{eq:int_schw}
\begin{align}
    E & \defeq -C(\xi_{(t)}) = -p_t = -P_t - \frac{M}{r^2}S^{tr} \:,\\
    \begin{split}
        J_x & \defeq C(\xi_{(x)}) = -\sin\phi \: p_\theta \\ &- \cos\phi \: \cot\theta \: p_\phi + r^2 \cos\phi \: S^{\theta\phi} = -\sin\phi\: P_\theta \\ &- \cos\phi\: \cot\theta\: P_\phi + r^2\cos\phi\: \sin^2\!\theta\: S^{\theta\phi} \\
        & + r\sin\phi\: S^{\theta r} + r\cos\phi\: \sin\theta\: \cos\theta\: S^{\phi r} \:, 
    \end{split}\\
    \begin{split}
        J_y &\defeq C(\xi_{(y)}) =  \cos\phi \: p_\theta \\ &- \sin\phi \: \cot\theta \: p_\phi + r^2 \sin\phi \: S^{\theta\phi} = \cos\phi\: P_\theta \\ &- \sin\phi\: \cot\theta\: P_\phi + r^2\sin\phi\: \sin^2\!\theta\: S^{\theta\phi}  \\
        & - r\cos\phi\: S^{\theta r} + r\sin\phi\: \sin\theta\: \cos\theta\: S^{\phi r} \:,
    \end{split}\\
    \begin{split}
        J_z &\defeq C(\xi_{(z)}) = p_\phi = P_\phi - r\sin^2\!\theta\: S^{\phi r}  \\
        & - r^2\sin\theta\: \cos\theta\: S^{\phi\theta} \:. 
    \end{split}
\end{align}
\end{subequations}
Using these, we define the measure of the total angular momentum as
\begin{equation}
    J^2 = J_x^2 + J_y^2 + J_z^2 \:,
\end{equation}
and by taking the Poisson bracket \eqref{eq:poisson} we can see that
\begin{subequations}\label{eq:involution}
\begin{align}
    \{E,J_j\} &= 0 \:,\\
    \{J_i,J_j\} &= -\epsilon_{ijk}J_k \:,\\
    \{J^2,J_j\} &= 0 \:.
\end{align}
\end{subequations}

The spin magnitude $S^2 = \sqrt{S_{\mu\nu}S^{\mu\nu}/2}$ and the expression $\epsilon_{\mu\nu\rho\sigma}S^{\mu\nu}S^{\rho\sigma} = 0$ are used  in the construction of the canonical formalism. They cannot be used to further reduce the DoF.

\subsubsection{Reduction of the system}

The system as described so far has 4 spacetime DoF and 1 spin DoF \cite{Witzany18}. It is possible to use the integrals of motion, i.e. $E$, $J_z$ and $J^2$, to reduce the DoF to only 2, as discussed in detail in Sec.~\ref{sec:integrability}. We have seen in Eq. \eqref{eq:involution} that they truly are in involution. This amounts to picking specific values of $E$, $J_z$ and $J^2$, which then become parameters of the reduced system.

In fact, we can go further: due to spherical symmetry of the Schwarzschild spacetime, for every orbit there exists a coordinate system such that the total angular momentum is aligned with the $z$-axis ($\theta = 0$) \cite{Suzuki96}. Thus, every relevant feature of the system's dynamics remains covered by making the choice
\begin{equation}
    J_x = J_y = 0 \:.
\end{equation}
Due to this, we can write
\begin{equation}
    \begin{split}
    \begin{pmatrix} 0 \\ 0 \end{pmatrix} &= 
    \begin{pmatrix} J_x \\ J_y \end{pmatrix} = \\
    &\begin{pmatrix} \cos\phi & -\sin\phi \\ \sin\phi & \cos\phi \end{pmatrix}
    \begin{pmatrix} -\cot\theta \: p_\phi + r^2 S^{\theta\phi} \\ p_\theta \end{pmatrix} \,,
    \end{split}
\end{equation}
and thus
\begin{equation}
    p_\theta = -\cot\theta \: p_\phi + r^2 S^{\theta\phi} = 0 \:.
\end{equation}

A useful tool to get an insight into the boundaries of the spinning particle motion would be an effective potential on the $r,~\theta$ plane. However, such an effective potential is not known, instead, as shown in \cite{Suzuki96}, there is a boundary curve on the $r,~\theta$ plane for given $E$, $J_z$ and $S$, on which $P_r = P_\theta = 0$. As is \cite{Suzuki96}, we will call this curve an "effective potential", even if it is not. This curve has two branches reading
\begin{subequations}
\begin{align}
    \begin{split}
    &V_\mathrm{eff\:\left(\pm\right)} = \mu\left[\sqrt{f}\cosh X_{\left(\pm\right)} \right. \\
    &\quad  \left.+ \frac{M\sinh X_{\left(\pm\right)}}{\sqrt{f}r\cosh X_{\left(\pm\right)}}\cdot\left(\frac{J_z\sin\theta}{\mu r} - \sinh X_{\left(\pm\right)}\right)\right] \:,
    \end{split} \\ 
    \begin{split}
    &\sinh X_{\left(\pm\right)} = \frac{\mu J_zr\sin\theta}{\denom} \pm \\ &\pm \frac{1}{\denom}\sqrt{\left(\mu J_z r\sin\theta\right)^2 - \mathcal{E}\denom} \:,
    \end{split} \\
    &\mathcal{E} = \left(J_z^2-S^2\right)f+\frac{2M}{r}J_z^2\sin^2\!\theta \:, \\
    &\denom = \mu^2 r^2 - S^2 f \:.
\end{align}
\end{subequations}

In the limit $r\to\infty$, non-negativity of the square root argument leads to the requirement
\begin{equation}
    \cos\theta \leq \frac{S}{J_z} \:,
\end{equation}
and for finite $r$, it seems to be even slightly more restrictive. This means that for low spin values the particle can only move in a thin wedge near the equatorial plane. In the limiting geodesic case $S\to 0$, the particle is confined to the equatorial plane (recall that we have imposed the constraint $J^2 = J_z^2$!).

\begin{figure}
    \centering
    \ifimages
     \includegraphics[width=0.45\textwidth]{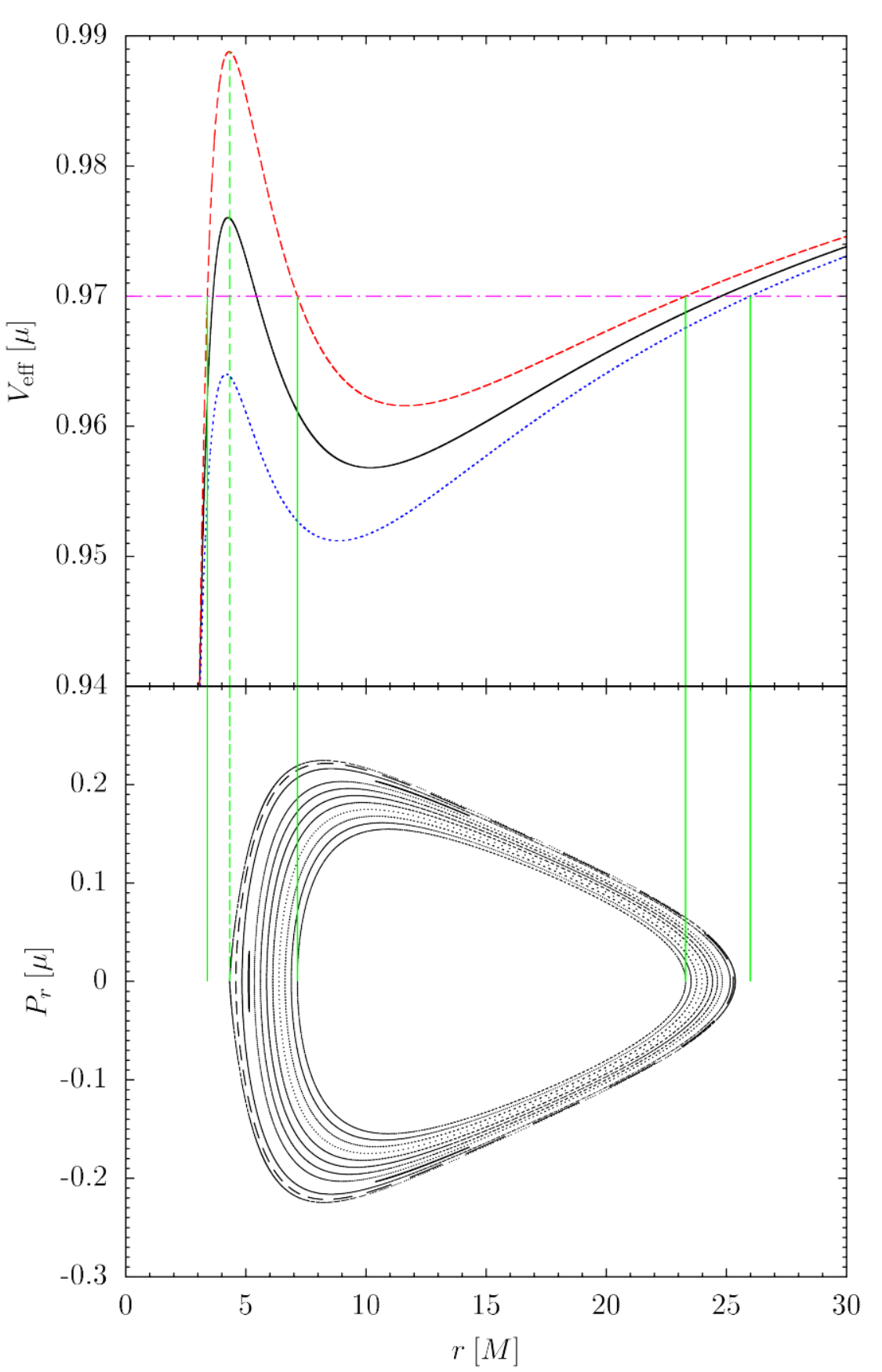}
    \fi
    \caption[Effective potential vs. Poincar\'{e} section]{Top panel: The effective potential in the equatorial plane for $J_z = 3.8\mu M$. The full black line corresponds to $S=0$, dashed red and dotted blue lines are $V_{\mathrm{eff}\left(+\right)}$ and $V_{\mathrm{eff}\left(-\right)}$, respectively, for $S=0.2\mu M$. The purple dash-and-dot line shows the energy value $E=0.97\mu$. Bottom panel: The corresponding Poincar\'{e} section \iffalse for $E=0.97\mu$, $J_z=3.8\mu M$, $S=0.2\mu M$ \fi with 12 initial conditions from $r=4.327M$ with spacing $0.256M$. The full green vertical lines connecting both plots show bounds for possible initial $r$ with $P_r = 0$ in the Poincar\'{e} section and the dashed green vertical line shows the approximate connection of the unstable orbit and the extrema of the effective potential.}
    \label{fig:mpd_veff1}
\end{figure}

For high values of the spin, we cannot make general statements about the effective potential; however, for $S \ll \mu M$, it will only deviate slightly from its geodesic counterpart. Typical shapes of $V_{\mathrm{eff}\left(\pm\right)}$ are shown in the top panel of Fig. \ref{fig:mpd_veff1}, where we can see the band of $r$ and $E$ values allowed as initial conditions when considering $P_r=0$ with the geodesic effective potential running between $V_{\mathrm{eff}\left(\pm\right)}$ curves. The bottom panel of Fig. \ref{fig:mpd_veff1} shows the corresponding Poincar\'{e} section for a given energy (the purple dash-and-dot line in the top panel). The full green vertical lines crossing both plots show the radial bounds defined by the effective potential, i.e. $P_r = 0$. The two inner green full lines define a region of non-existing initial conditions on the Poincar\'{e} section. One can see that in the outer part of the surface of section still within the band of admissible initial conditions (outer green full lines), there are no points: the corresponding initial conditions exist but correspond to fast plunging orbits most of which only intersect the surface of section very few times before plunging into the central object and therefore do not form structures in the figure. Thus, for a given energy both branches of the effective potential are needed to define the bounds of the allowed motion. Note that this holds independently from the relevant orientation of the spin and the angular momentum.\footnote{The relevance between the effective and the bounds of the allowed motion is not accurately described in \cite{Suzuki96}.}

\subsubsection{Initial conditions}

In this section, we describe the choice of initial conditions for numerical integration of orbits. To gain insight into the phase space dynamics, we reduce the system to only 2 degrees of freedom as described in the previous section. In particular, for every orbit, we will choose the values of the energy $E$, total angular momentum $J = \sqrt{J^2}$ and spin $S = \sqrt{S^2}$. 

Since the motivation for this work are EMRI, it is necessary to pick values of the spin which are astrophysically relevant. The extended body we are modeling can be either a 
\begin{itemize}
    \item stellar mass black hole, in which case calculations with the Kerr spacetime show that $S\leq \mu^2$, or
    \item neutron star, wherein due to the mass shedding limit $S\lesssim 0.6\mu^2$ (see, e.g., \cite{Hartl02}).
\end{itemize}
The importance of the spin term in the MPD equations in Schwarzschild spacetime scales as $S/(\mu M)$, and $S$ must in either case must be bounded by $\mu^2$. The strength of the astrophysically realistic perturbation to the equations of motion of the particle thus is of the order of magnitude
\begin{equation}
    \frac{S}{\mu M} \leq \frac{\mu^2}{\mu M} = \frac{\mu}{M} \lesssim 10^{-4} \:,
\end{equation}
where we use $10^{-4}$ as an upper bound for the mass ratio $\mu/M$ in an EMRI.

As a surface of section, we choose the equatorial plane $\theta=\pi/2$ with the momentum pointing "down", i.e. $P_\theta \geq 0$. As coordinates on the section, we choose the radial coordinate and momentum $r, P_r$. This is a very usual choice in black hole spacetimes, employed in previous works (see, e.g., \cite{LukesGerakopoulos12,LukesGerakopoulos17b} and references therein).

\begin{figure}
    \centering
    \ifimages
    \includegraphics[width=0.45\textwidth]{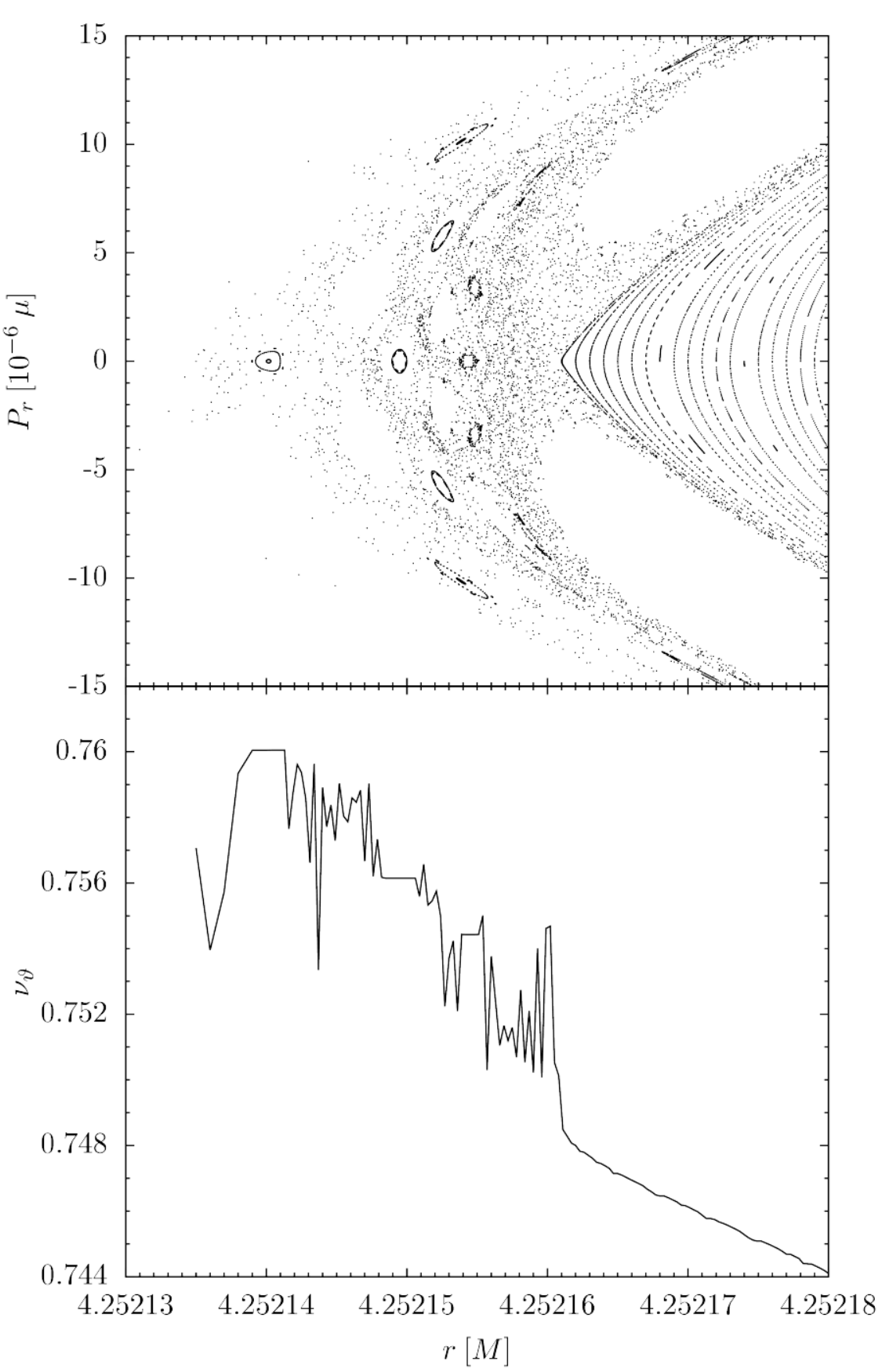}
    \fi
    \caption[Poincar\'{e} section and rotation curve for $S=10^{-4}\: \mu M$]{Top panel: Poincar\'{e} section of the MPD equations for $S=10^{-4}\: \mu M$, $E=0.976\mu$, $J_z=3.8\mu M$ and initial $P_r = 0$. Initial $r$ are 31 equidistant values from $4.252135M$ with the spacing $10^{-6}\: M$. Bottom panel: rotation curve corresponding to the top panel.}
    \label{fig:mpd_sec2}
\end{figure}

The top part of Fig.~\ref{fig:mpd_sec2} shows the left tip of a Poincar\'{e} section and corresponding rotation curve for spin value $S=10^{-4}\: \mu M$. Visible are all the typical features of a non-integrable system: KAM curves on the right, small islands of stability as remnants of resonant tori and dispersed points corresponding to chaotic orbits (from the center to the left). The bottom part of Fig.~\ref{fig:mpd_sec2} shows the corresponding rotation curve with initial conditions taken at $P_r = 0$. Note that the rotation curve corresponds to the form described in Sec.~\ref{sec:perturbation} it forms a plateau when passing through a secondary island of stability, it is monotonous when passing through KAM tori in the main island of stability, and it varies wildly in the chaotic sea. Therefore, the MPD equations with the TD SSC exhibit very typical signs of a chaotic system even for a small enough spin value to correspond to an EMRI with $\mu/M \doteq 10^{-4}$. 

\section{Action Angle(-like) variables}\label{sec:aa}

\subsection{Action Angle variables}

As mentioned in the previous section, in an integrable system, there exist canonical coordinates in the form of Eq. \eqref{eq:lin_variables}. Their existence is connected to the separability of the Hamilton-Jacobi equation into a set of independent ordinary differential equations by assuming the action
\begin{equation}
    \mathcal{W} = \sum_{i=1}^N  \mathcal{W}_i\left(q^i\right) \:.
\end{equation}
If the Hamilton-Jacobi equation is separable in the given variables, the actions are defined as
\begin{equation}
    I_i = \oint \der{\mathcal{W}_i}{q^i} \d q^i \:,
\end{equation}
where the integral is taken over one loop of the corresponding variable. The angles are then such variables that they fulfill the canonical Poisson bracket
\begin{equation}
    \{\theta^i,I_j\} = \delta^i_j
\end{equation}
and are $2\pi$-periodic. The $I_i$ are integrals of motion, i.e.
\begin{equation}
    \pder{H}{\theta^i} = 0 \:.
\end{equation}
The values of the actions determine on which phase space torus the motion lies and the angles determine the position on the given torus. This set of coordinates is then called the Action-Angle (AA) variables.

\subsection{Growth of resonances}\label{sec:gro_theory}

\begin{figure}
    \centering
    \ifimages
     \includegraphics[width=0.45\textwidth]{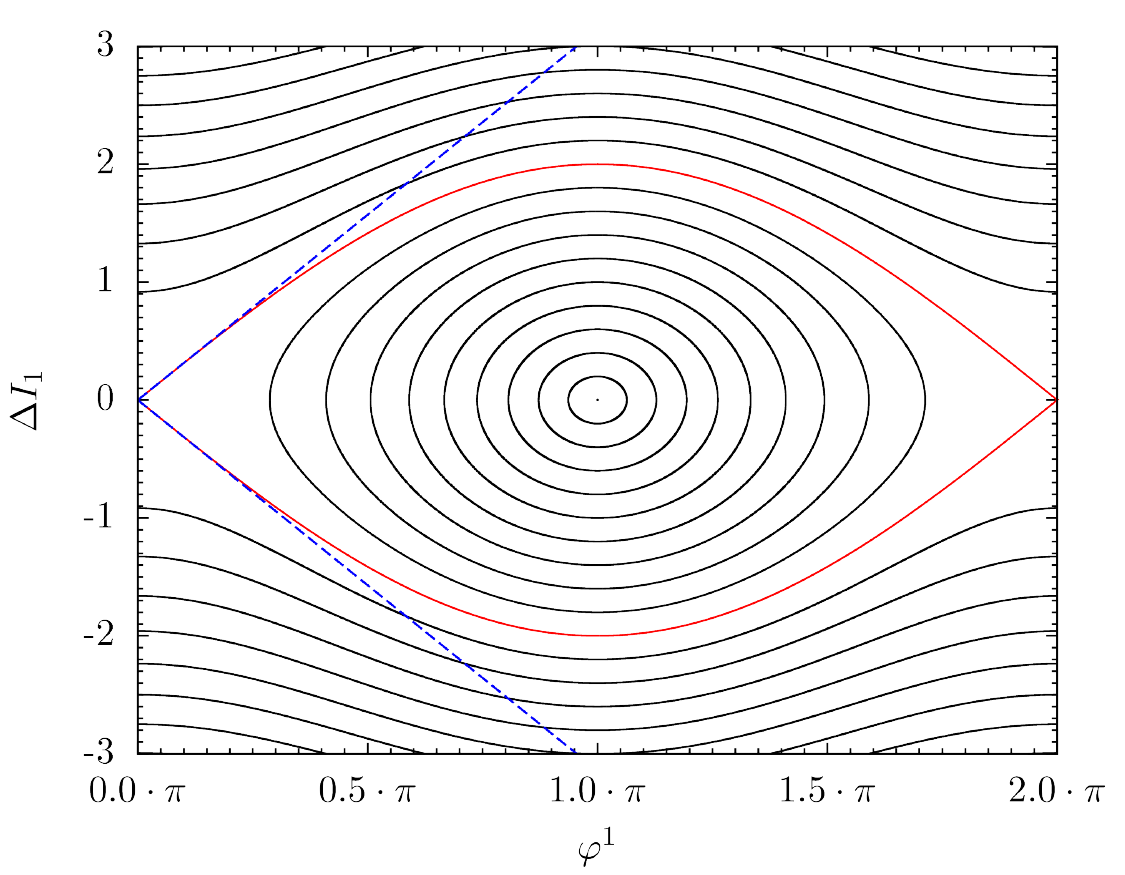}
    \fi
    \caption[Separatrix and island of stability of the pendulum model]{Phase portrait of the non-linear oscillator of Eq. \eqref{eq:osc_phase1} for $n=1$ and $\beta = \varepsilon\alpha = 1$. The full curve in red is the separatrix, the angle at which it opens is measured between the dashed blue lines and the full black lines correspond to trajectories of the system; the ones enclosed between the two separatrix branches correspond to oscillation and the outer ones to libration.}
    \label{fig:osc_phase1}
\end{figure}

In a non-inteegrable system, AA variables cannot be defined for the whole system. Nevertheless, if one defines a set of variables which smoothly reduce to AA variables when the perturbation is eliminated, it is possible to study the effect of gradual deviation from the integrable case using perturbation theory and series expansions. We have written this section along the lines of \cite{Morbidelli02}, for more reference, we also recommend \cite{Lichtenberg92}, \cite{Arnold06}.

We assume an integrable Hamiltonian system with 2 DoF in AA variables $\theta^i, I_i, i=1,2$ with Hamiltonian $H = H_0\left(I_1, I_2\right)$. To focus on the resonance with rotation number $\omega = r/s$, we define the angle $\varphi^1 \defeq s\theta^1 - r\theta^2$, which, in the integrable case, would be zero. To describe the deviation from the unperturbed system, the Hamiltonian to first order in the perturbation parameter $\varepsilon$ can then be written as \cite{Morbidelli02}
\begin{equation}
    H = H_0\left(I_1, I_2\right) + \varepsilon H_{1}\left(\varphi^1, I_1, I_2\right).
\end{equation}
Also expanding the actions $I_i = I_i^0 + \Delta I_i$, one gets to second order in $\Delta I_1$
\begin{equation}
    H = \frac{\beta}{2}\left(\Delta I_1\right)^2 + \varepsilon F\left(\varphi^1\right) \:,
\end{equation}
where we have omitted the $I_i^0$ dependence and the constant term. The last step is to expand the $F$ in a Fourier series and only keep the main term, arriving, without loss of generality, at
\begin{equation}\label{eq:osc_phase1}
    H = \frac{\beta}{2}\left(\Delta I_1\right)^2 + \varepsilon \alpha\cos\left(n\varphi^1\right) \:.
\end{equation}

For $n=1$, this is the Hamiltonian of a pendulum, revealing a certain universality of resonances. The separatrix is the curve of constant Hamiltonian, separating the phase space regions corresponding to oscillation and libration: $\varepsilon\alpha = \beta\left(\Delta I_1\right)^2/2 + \varepsilon\alpha\cos\left(n\varphi^1\right)$. The minima of $\cos\left(n\varphi^1\right)$ with $\Delta I_1 = 0$ form stable points and the maxima form unstable points. A straightforward calculation then shows that the width of the resonance is
\begin{equation}\label{eq:pert_to_wid}
    \mathrm{width} \defeq \max\left(I_1\right) - \min\left(I_1\right) = 4\sqrt{\frac{\varepsilon\alpha}{\beta}} \propto \sqrt{\varepsilon} \:.
\end{equation}
This model also allows us to relate the width to the angle at which the separatrix opens at an unstable point. In a neighborhood of $\varphi^1 = 0$, one may write for the separatrix
\begin{equation}
    \Delta I_1 = \pm \sqrt{\frac{\epsilon\alpha}{\beta}}\cdot n\varphi^1 = \pm\frac{\mathrm{width}\cdot n}{4}\cdot\varphi^1 \:.
\end{equation}
Now it is only necessary to return to the original variables $\theta^i, I_i, i=1,2$. For a given value of $\theta^2$ - without loss of generality, we can take $\theta^2 = 0$:
\begin{equation}
    \theta^1 = \frac{\varphi^1}{s} \:,
\end{equation}
revealing that in the Poincar\'{e} section $\theta^2 = 0$, there will be a total of $n\cdot s$ islands in the given resonance. This way we can see that in the weakly perturbed system the $n$ in the discussion above corresponds to the $n$ given in the Poincar\'{e}-Birkhoff theorem in Sec.~\ref{sec:perturbation}. This also allows us to finally write the relationship between the resonance width and the local behavior of the separatrix:
\begin{equation}\label{eq:ang_to_wid}
    \Delta I_1 = \pm \frac{\mathrm{width}\cdot ns}{4}\cdot \Delta\theta^1 \:.
\end{equation}

Thus, to determine the width of a resonance in a non-linear system, we can measure the angle at which the separatrix opens - that is, between the Jacobian's eigenvectors at the corresponding unstable periodic point - in AA variables and make use of Eq. \eqref{eq:ang_to_wid}. We can also employ the non-linear oscillator model to estimate the rate at which resonances of a spinning particle near a black hole grow. It has been shown that for MPD equations with the TD SSC to linear order in spin in the Kerr spacetime (of which Schwarzschild is a special case) approximate constants of motion exist that allow for an (approximate) separation of the Hamilton-Jacobi equation (see \cite{Ruediger81, Witzany19}). It is not clear whether this means that no resonances appear at linear-in-spin order.

As shown above, for a perturbation linear in the parameter $\varepsilon$, resonances grow as $\mathrm{width}\propto \sqrt{\varepsilon}$. If the terms linear-in-spin in the equations of motion cause a given resonance to appear, then one can expect the resonance to grow as $\mathrm{width}\propto \sqrt{S}$. However, as it is possible that the resonance only appears due to terms which are second order in spin, then we must equate the parameter $\varepsilon$ to $S^2$ and get $\mathrm{width} \propto \sqrt{S^2} = S$.

Hence, there are two cases that may appear: $\mathrm{width} \propto \sqrt{S}$ or $\mathrm{width} \propto S$.

\subsection{AA-like variables for the MPD equations}

\begin{figure}
    \centering
    \ifimages
     \includegraphics[width=0.45\textwidth]{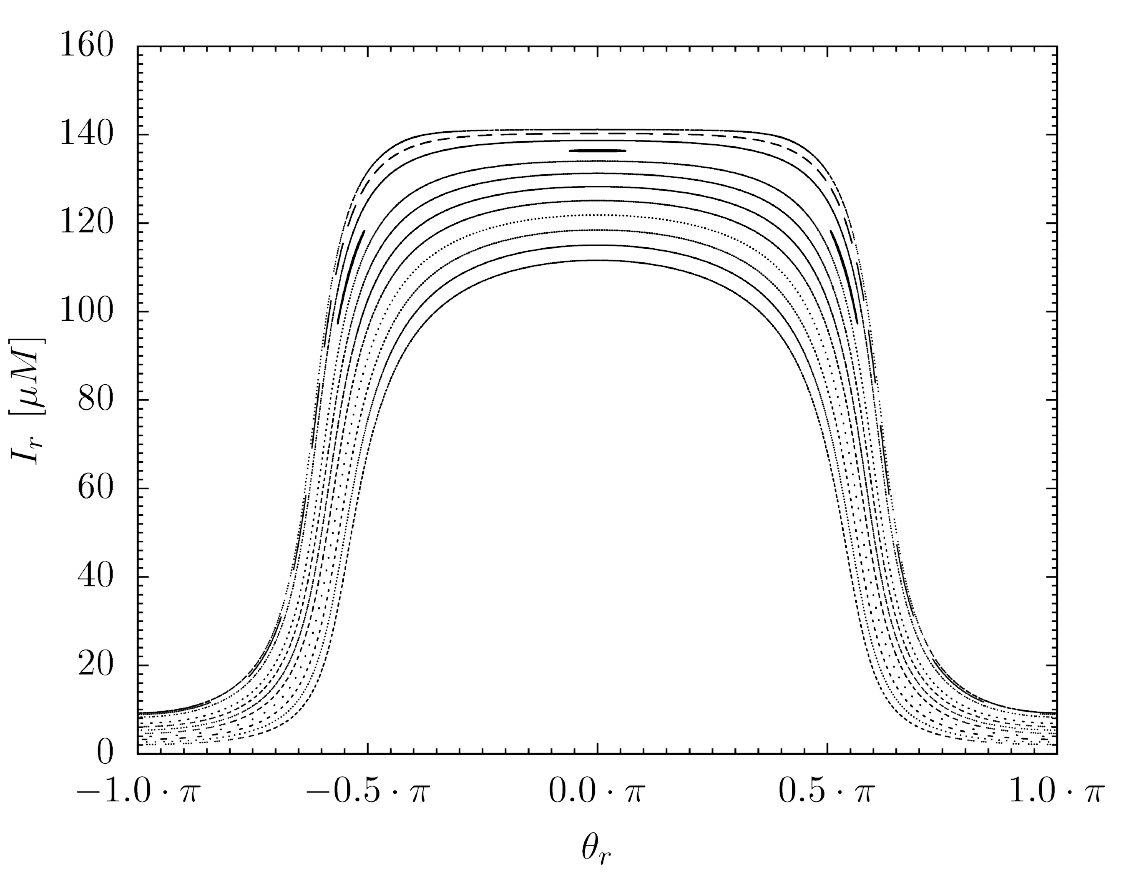}\\
     \includegraphics[width=0.45\textwidth]{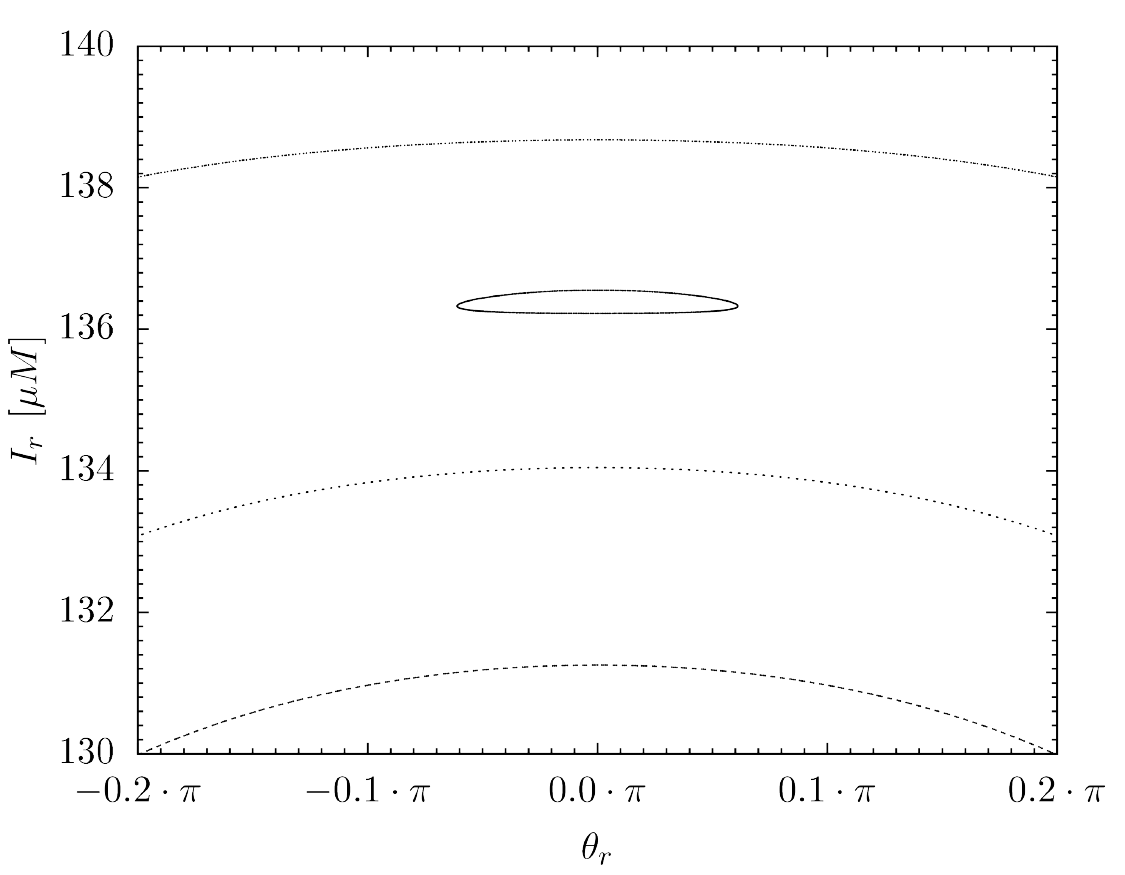}
    \fi
    \caption[AA-like variables]{Data from the surface of section in Fig. \ref{fig:mpd_veff1} converted to AA-like variables. Top panel: full surface of section, bottom panel: zoom at the $1/3$ island.}
    \label{fig:mpd_aa1}
\end{figure}

To find a reasonable mapping between our $r,P_r$ surface of section with given $E$, $J_z$, $J_x=J_y=0$, $S$ and a set of AA-like variables, we make use of the geodesic properties: we find a suitable set of fiducial parameters $E_f$, $J_{zf}$ and use the geodesic $\left(r,P_r\right) \leftrightarrow \left(\theta^r,I_r\right)$ mapping with these fiducial parameters to convert our Poincar\'{e} section data to AA-like variables.

Our approach of finding suitable values for the fiducial parameters relies on the existence of an unstable periodic point $\left(r_{\mathrm{upo}},0\right)$ near the left tip of the surface of section (e.g., Fig.~\ref{fig:mpd_sec2}); our method preserves this unstable point when mapping to the geodesic system. The fiducial parameters are then given by the geodesic formulae
\begin{align}
    E_f &= \frac{r_{\mathrm{upo}}-2M}{\sqrt{r_{\mathrm{upo}}\left(r_{\mathrm{upo}}-3M\right)}}\mu \:, \\
    J_{zf} &= \frac{r_{\mathrm{upo}}}{\sqrt{M\left(r_{\mathrm{upo}}-3M\right)}}\mu M \:.
\end{align}

Once the $E_f$ and $J_{zf}$ are known, we convert the point $\left(r, P_r\right)$ in the Poincar\'{e} section to $\left(\theta^r, I_r\right)$ as a point on a geodesic with energy $E_f$, azimuthal angular momentum $J_{zf}$, radial coordinate $r$ and corresponding momentum $P_r$ passing through the equatorial plane:
\begin{subequations}
\begin{align}
    \begin{split}
    I_r = I_r^{geo}&\left(r;E=E_f,J_z=J_{zf}, \right. \\
    &\left. \mathcal{C}=r^2\left[\frac{E^2}{f}-fP_r^2-\mu^2\right]-J_z^2\right) \:, 
    \end{split} \\
    \begin{split}
    \theta^r = \theta^r_{geo}&\left(r;E=E_f,J_z=J_{zf}, \right. \\
    &\left. \mathcal{C}=r^2\left[\frac{E^2}{f}-fP_r^2-\mu^2\right]-J_z^2\right) \:.
    \end{split}
\end{align}
\end{subequations}

As the choice of fiducial parameters $E_f$, $J_{zf}$ is made using the left tip of the surface of section, the transformation is expected to work well for $r \sim r_{\mathrm{upo}} \Leftrightarrow \theta^r \sim 0$ and might be unsuitable for $r \gg r_{\mathrm{upo}} \Leftrightarrow \abs{\theta^r} \sim \pi$. This is demonstrated in Fig. \ref{fig:mpd_aa1}, where the previously shown surface of section in Fig. \ref{fig:mpd_veff1} is converted to AA-like variables and one can see that for high $I_r$ and low $\abs{\theta^r}$ they behave as expected, i.e. forming approximately horizontal lines for KAM circles.

\subsection{Numerical analysis of resonance growth}

\begin{figure}
    \centering
    \ifimages
     \includegraphics[width=0.45\textwidth]{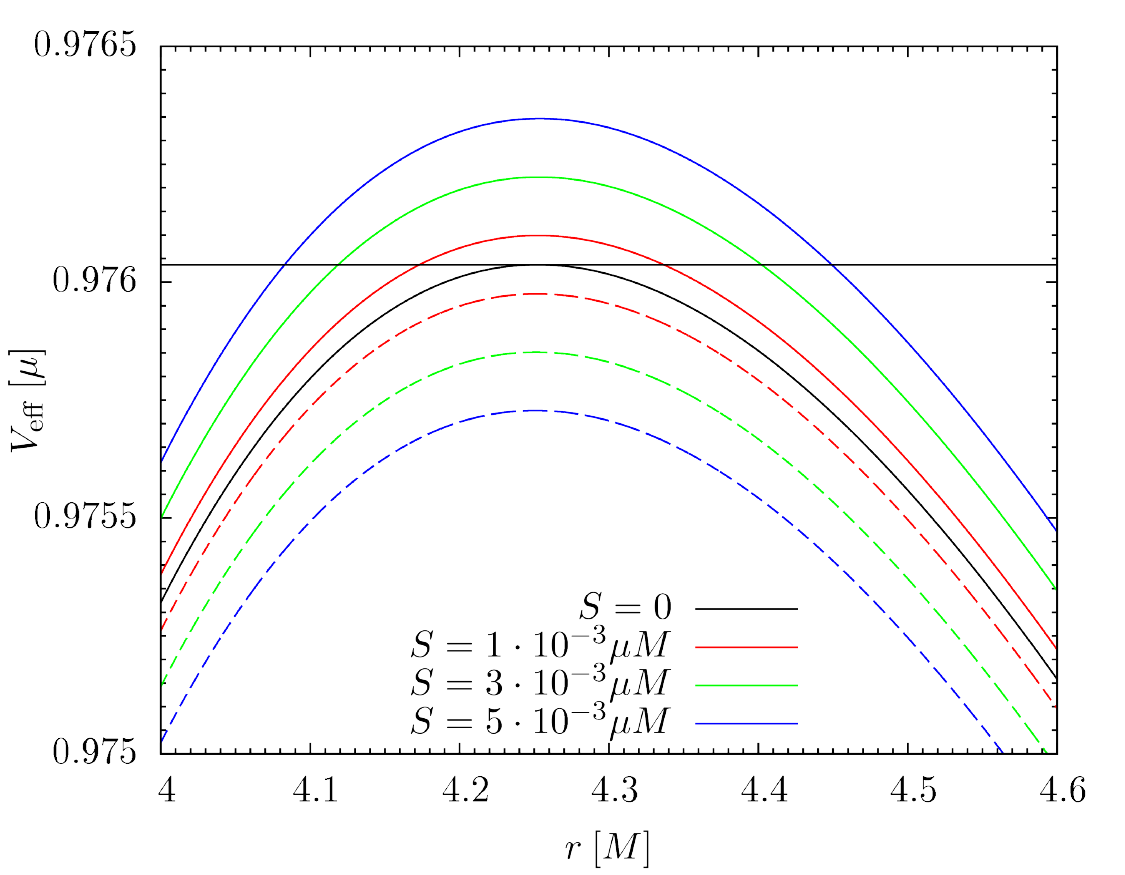}
    \fi
    \caption[Effective potential]{Shapes of the effective potential for angular momentum $J_z=3.8\mu M$ andq low spin values near the unstable point. Black lines show the geodesic effective potential and the corresponding energy level of the circular equatorial unstable orbit at $E \doteq 0.976\:037\mu$. Coloured lines show the effective potential for non-zero spin; full lines correspond to $V_{\mathrm{eff}\left(+\right)}$ and dashed lines to $V_{\mathrm{eff}\left(-\right)}$.}
    \label{fig:mpd_veff2}
\end{figure}

\begin{figure}
    \centering
    \ifimages
      \includegraphics[width=0.45\textwidth]{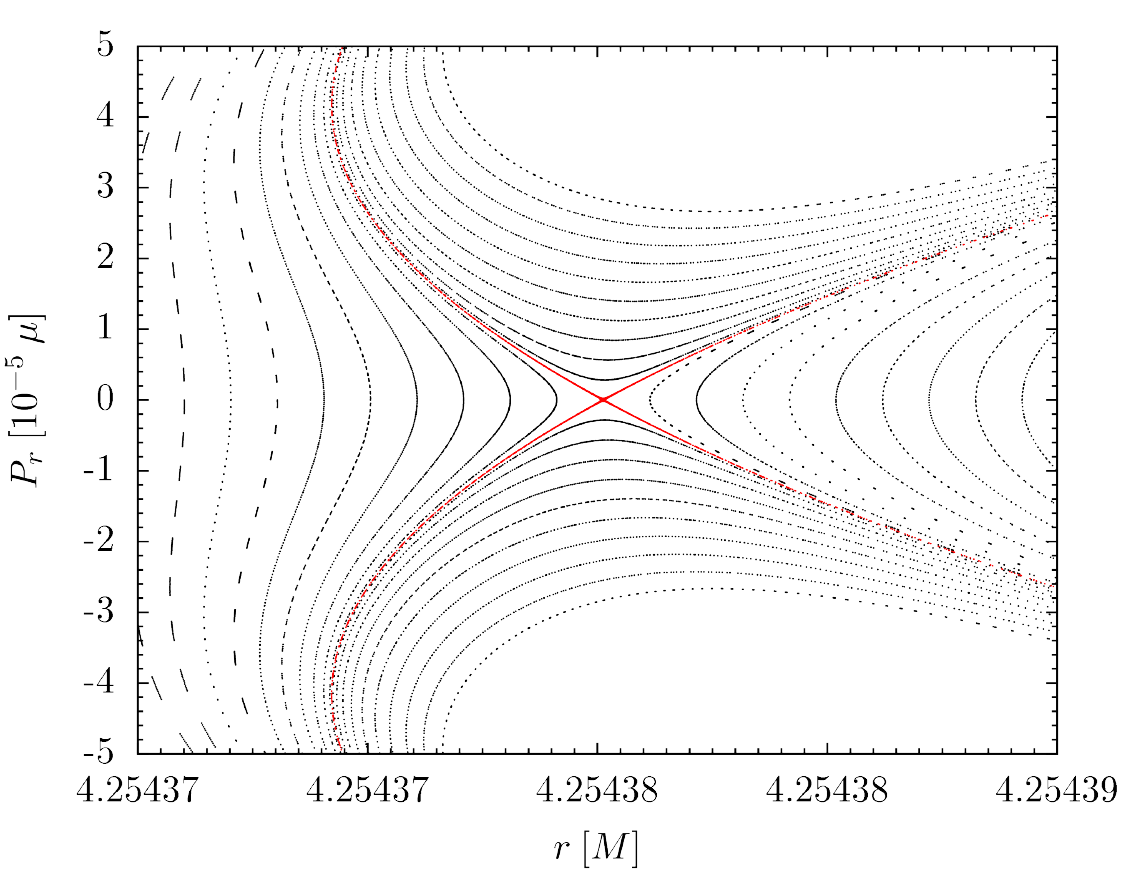}\\
      \includegraphics[width=0.45\textwidth]{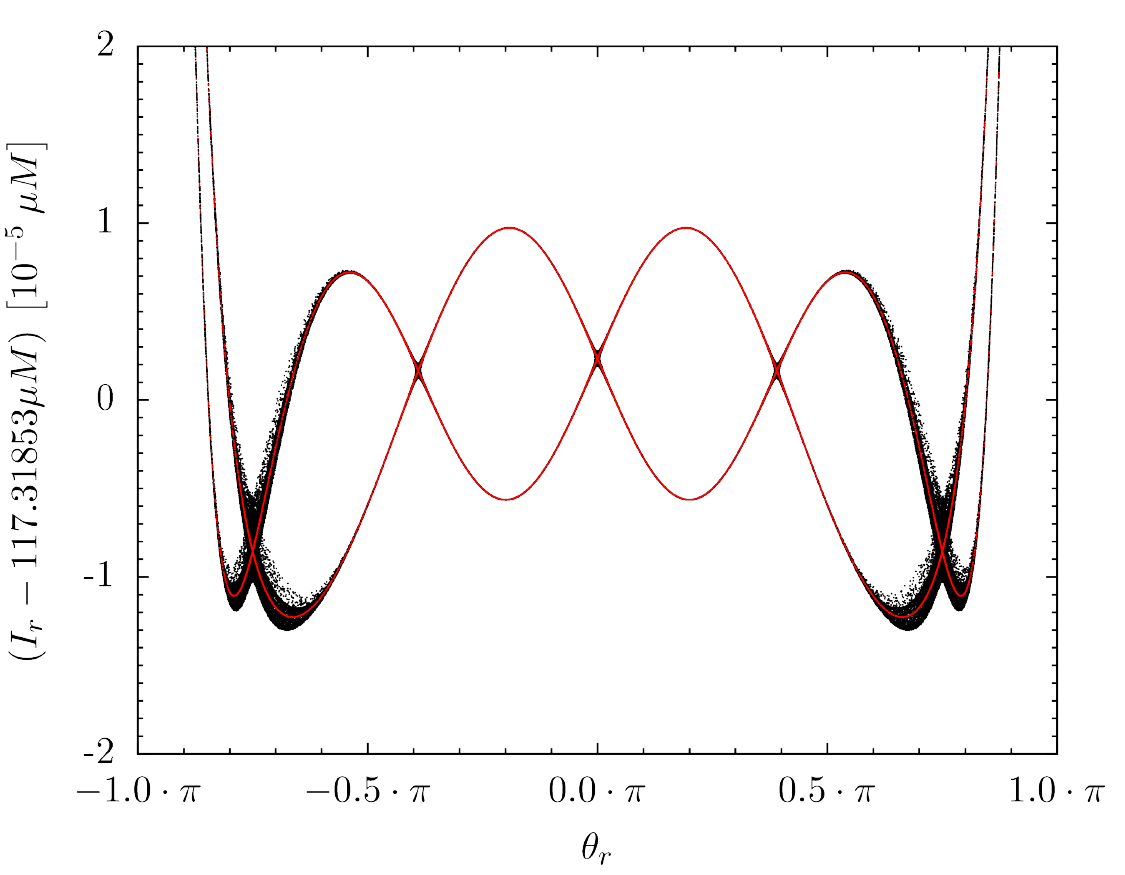}
    \fi
    \caption[Resonance $2/3$]{Top panel: Surface of section for $E\doteq 0.976\:037\mu$, $J_z = 3.8\mu M$, $S=10^{-3}\: \mu M$ showing the $2/3$ resonance; separatrix is shown in red, black lines left and right of the separatrix are KAM curves, above and under are in the resonant islands of stability. Bottom panel: top panel converted to AA-like variables.}
    \label{fig:mpd_sec3}
\end{figure}

The aim of this section is to describe the method and results of our numerical analysis. To study the growth of resonances, one must first design a scheme to choose the values of the integrals for different spins. A good guide is again the effective potential. Our goal is that the energy and angular momentum allow for plunging orbits, as this is suggested by some works (see, e.g., \cite{Contopoulos12}) to make non-linear behavior more prominent even for a resonance enclosed by an KAM circle; also, we defined the AA-like variables in such a way that we require the existence of an unstable point near the left tip of the main island of stability. To fulfill this criterion for small enough spins ($S\lesssim 10^{-1}\: \mu M$), we simply choose a value for $J_z$ and take the value of the geodesic effective potential at its local maximum (corresponding to an unstable circular equatorial geodesic orbit) as the energy, which we keep the same for different spin values. Typically, the effective potentials then looks as in Fig. \ref{fig:mpd_veff2}. We then use a program (see Appendix~\ref{sec:mpd_code}) to search for the given resonance along the $r=0$ line and after locating the separatrix, the angle at which it opens is measured and the resonance width is calculated.

\begin{figure}
    \centering
    \ifimages
     \includegraphics[width=0.45\textwidth]{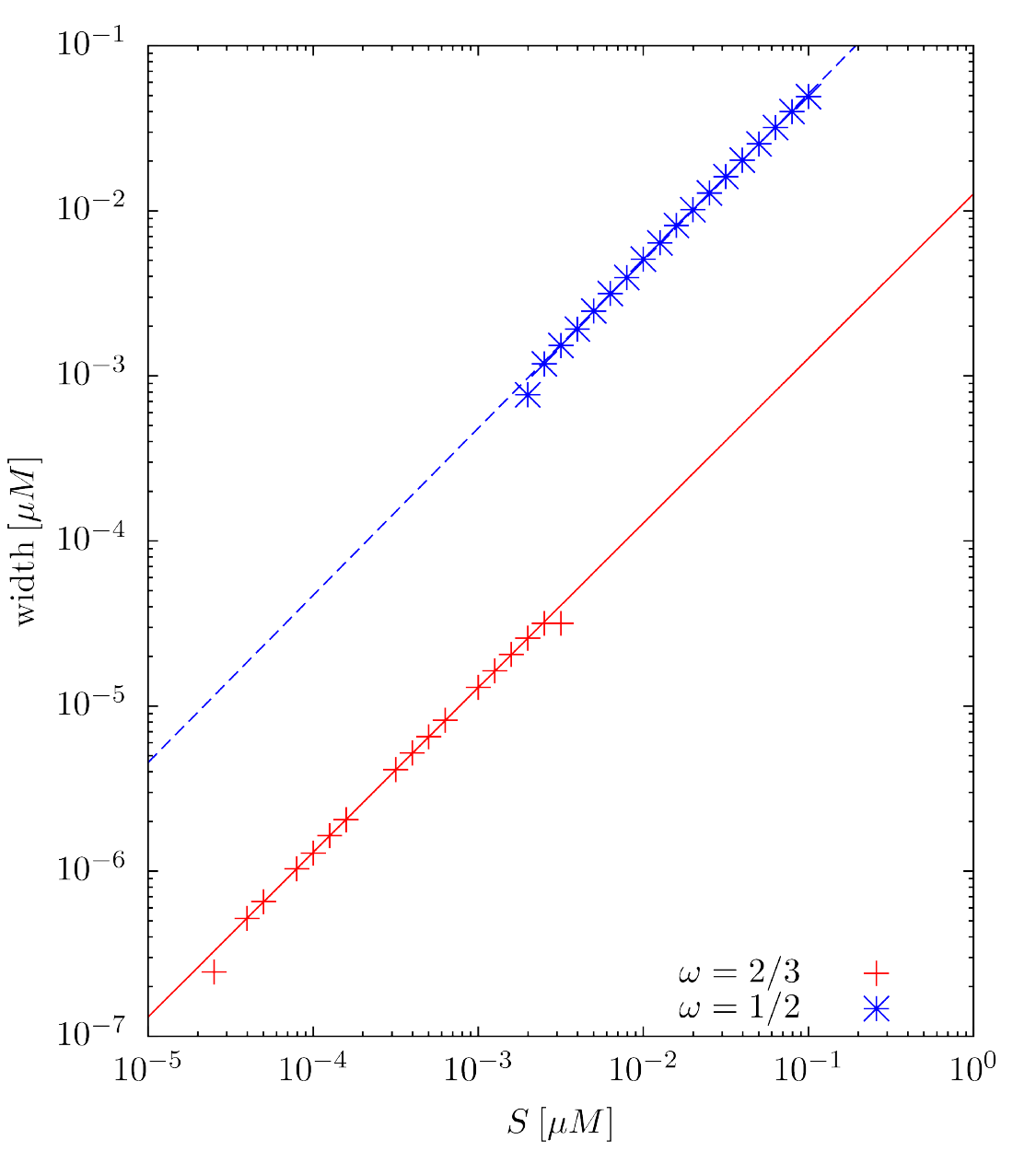}
    \fi
    \caption{Growth of resonances $\omega=1/2, 2/3$.}
    \label{fig:mpd_gro1}
\end{figure}

The growth of the $\omega=1/2$ and $2/3$ resonances with $J_z = 3.8\mu M$ was investigated by taking 21 different values of the spin distributed geometrically in the interval $\left[10^{-3}\: \mu M, 10^{-1}\: \mu M\right]$ for $\omega=1/2$ and 31 values in $\left[10^{-5}\: \mu M, 10^{-2}\: \mu M\right]$ for $\omega=2/3$; then the procedure above was applied to measure the width of the given resonance near the left tip of the main island of stability. Eq. \eqref{eq:ang_to_wid} was used to estimate the width of the resonance with parameter $n = 2$. The result is shown in Fig. \ref{fig:mpd_gro1}, where the lines correspond to fits of the function
\begin{equation}
    \log \frac{\mathrm{width_\omega}}{\mu M} = A_\omega + q_\omega\cdot\log\frac{S}{\mu M} \:,
\end{equation}
to the measured points except for the leftmost and rightmost ones, which was performed in logscale without taking errors of separatrix width measurement into account and returned the values
\begin{subequations}
\begin{alignat}{3}
    A_{1/2} &= -&&0.639  &&\pm 0.019 \:, \\
    q_{1/2} &=  &&1.013  &&\pm 0.004 \:, \\
    A_{2/3} &= -&&4.366  &&\pm 0.013 \:, \\
    q_{2/3} &=  &&0.9974 &&\pm 0.0016 \:.
\end{alignat}
\end{subequations}
This shows that the resonance widths grow linearly with spin. Recall that the resonance grows as the square root of the relevant perturbation (see Sec. \ref{sec:gro_theory}); we can thus conclude that the growth of the 1/2 and 2/3 resonances is driven by the second-order in spin terms.

\section{Gravitational waveforms}\label{sec:waves}

\subsection{Teukolsky equation}

The Teukolsky equation employs the Newmann-Penrose (NP) formalism to describe perturbations of test fields in a Kerr background. We provide only a short summary, the interested reader is referred to \cite{Teukolsky73, Harms14}.

In a suitable tetrad, it holds that the Weyl-NP scalar
\begin{equation}
\psi_4 \defeq C_{\mu\nu\rho\sigma} n^\mu \bar{m}^\nu n^\rho \bar{m}^\sigma = \frac{\ddot{h}_+ - i\ddot{h}_\times}{2} \:.
\end{equation}
For the master variable $\Psi = r^{-4}\psi_4$ and spin weight $s=-2$, gravitational perturbations of the Schwarzschild spacetime (see \cite{Teukolsky73, Harms14} for Kerr) are described by the master equation
\begin{align}\label{eq:master}
    \begin{split}
    & \frac{r^2}{f}\pder[2]{\Psi}{t} - \frac{1}{\sin^2\!\theta}\pder[2]{\Psi}{\phi} 
    - \frac{1}{\sin\theta}\pder{}{\theta}\left(\sin\theta\pder{\Psi}{\theta}\right) \\
    &- \left(\frac{1}{r^2f}\right)^s\pder{}{r}\left[\left(r^2f\right)^{s+1}\pder{\Psi}{r}\right] 
    -2s\frac{i\cos\theta}{\sin^2\!\theta}\pder{\Psi}{\phi} \\
    &  - 2s\left(\frac{M}{f}-r\right)\pder{\Psi}{t} + s\left(s\cot^2\theta-1\right)\Psi = 4\pi r^2 T \:.
    \end{split}
\end{align}
The source term $T$ is then a complicated linear combination of the stress-energy tensor components corresponding to the tetrad vectors $n$ and $\bar{m}$.

The conventional way to solve this equation is in the frequency domain, in which it is fully separable (see, e.g, \cite{Press73}). This approach is highly effective for a geodesic source, because it is sufficient to work with very few frequencies. In the case of more complicated dynamics, however, a time-domain approach is preferable. Due to the spacetime axisymmetry, the $\phi$ degree of freedom is still separable.

Since we are interested in the gravitational radiation at distances very large in comparison to $M$, the quantity of interest will be the strain $h$ at null infinity $\mathcal{J}^+$. It can be decomposed as
\begin{equation} \label{eq:straindecom}
    h = \sum_{\ell=2}^\infty \sum_{\mathrm{m}=1}^\ell h_{\ell\mathrm{m}}\cdot\: _ {-2}Y_{\ell\mathrm{m}}\left(\theta,\phi\right) \:,
\end{equation}
where $_{s}Y_{\ell\mathrm{m}}\left(\theta,\phi\right)$ are the spin-weighted spherical harmonics with spin-weight $s$.

\subsection{Results: generated waveforms}\label{ssec:waveforms}

In this section, waveforms from trajectories depicted in section \ref{sec:mpd} are computed using the Teukode (see Appendix \ref{sec:teukode} or \cite{Harms14, Harms16}).

\begin{figure}
    \centering
    \ifimages
     \includegraphics[width=0.45\textwidth]{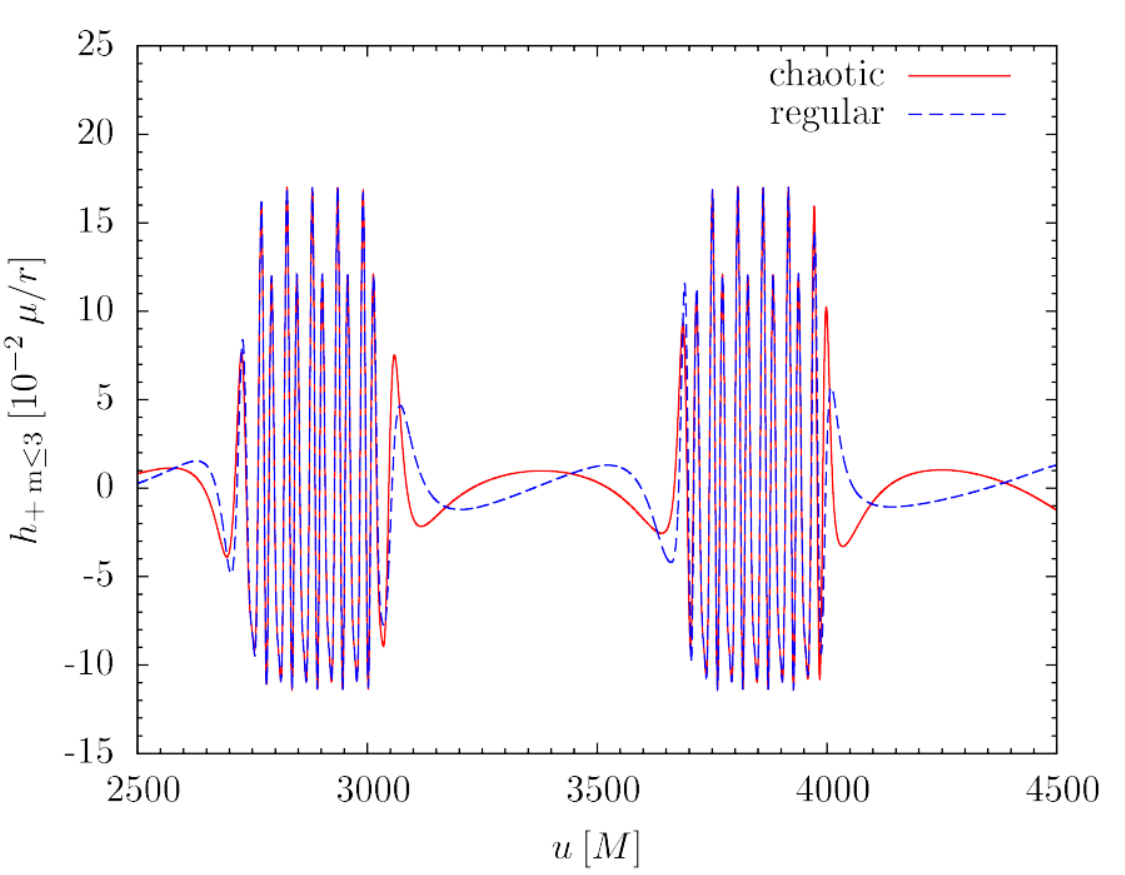}
    \fi
    \caption[Waveforms for $S=10^{-4}\: \mu M$]{Waveforms from trajectories in Fig.~\ref{fig:mpd_sec2}. In solid red: chaotic with initial $r = 4.252160M$ and $P_r = 0$, in dashed blue: regular with initial $r = 4.252162M$ and $P_r = 0$.}
    \label{fig:mpd_wav2}
\end{figure}

The waveforms shown are evaluated at null infinity $\mathcal{J}^+$ in the equatorial plane $\theta = \pi/2$. Fig.~\ref{fig:mpd_wav2} shows a waveform computed using $2001\times 101$ grid from a trajectory in the chaotic sea in Fig. \ref{fig:mpd_sec2} and a waveform from a nearby regular orbit. The orbits have eccentricity $e \approx \left(r_{\mathrm{max}}-r_{\mathrm{min}}\right)/\left(r_{\mathrm{max}}+r_{\mathrm{min}}\right) \doteq 0.776$ and get very close to the unstable orbit: $r_{\mathrm{min}}^{\mathrm{chaotic}} - r_{\mathrm{upo}} \doteq 5.3\cdot 10^{-5}\: M$, $r_{\mathrm{min}}^{\mathrm{regular}} - r_{\mathrm{upo}} \doteq 5.7\cdot 10^{-5}\: M$. These waveforms have the shape we expect from an EMRI (see, e.g., \cite{Drasco05}).

We can see that a simple visual inspection is insufficient to distinguish whether the original orbit was regular or chaotic. We propose and apply a solution in the next section.

\section{Recurrence analysis}\label{sec:rps}

\begin{figure*}
    \centering
    \ifimages
    {\includegraphics[width=0.3\textwidth]{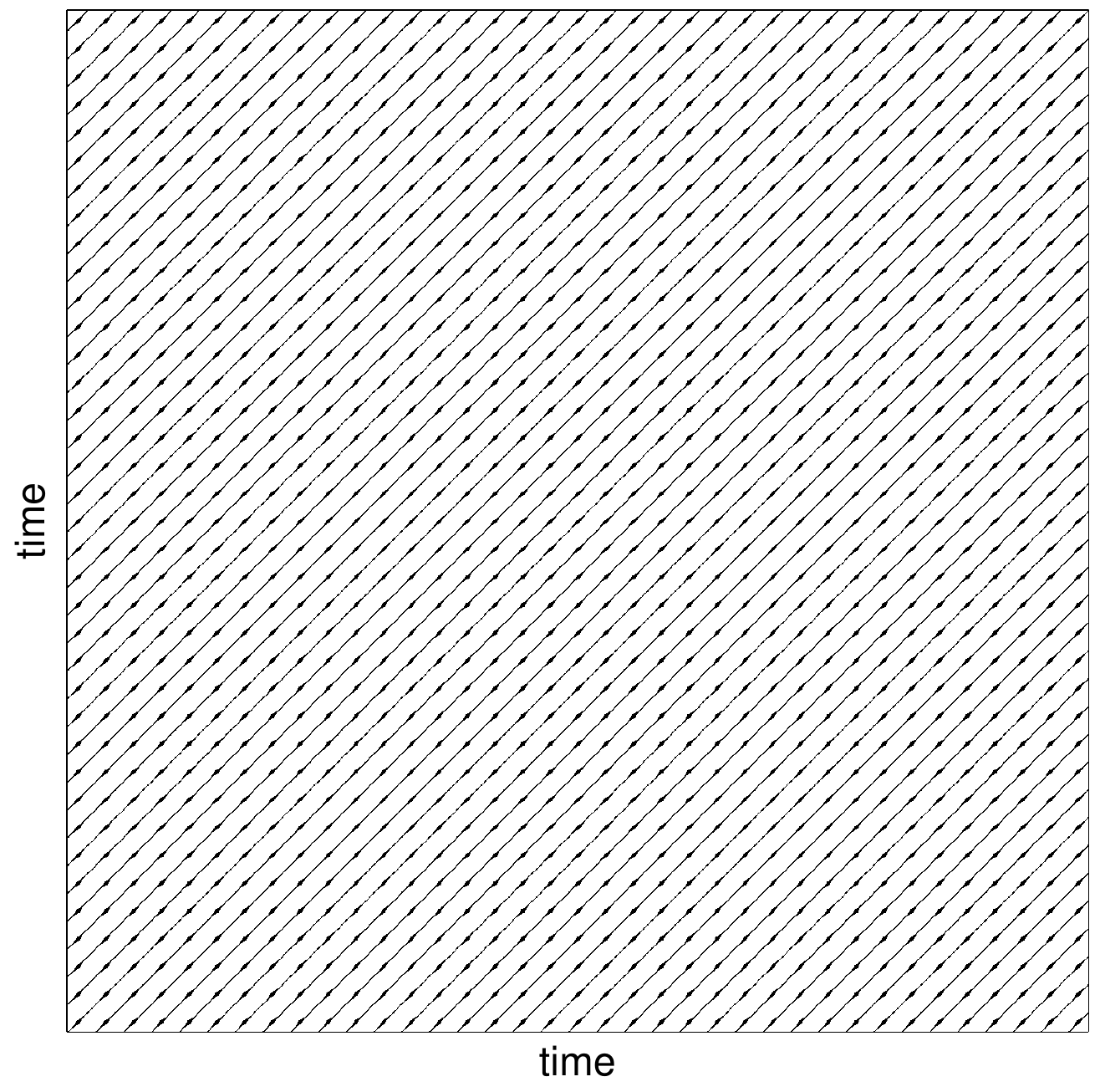}
    \includegraphics[width=0.3\textwidth]{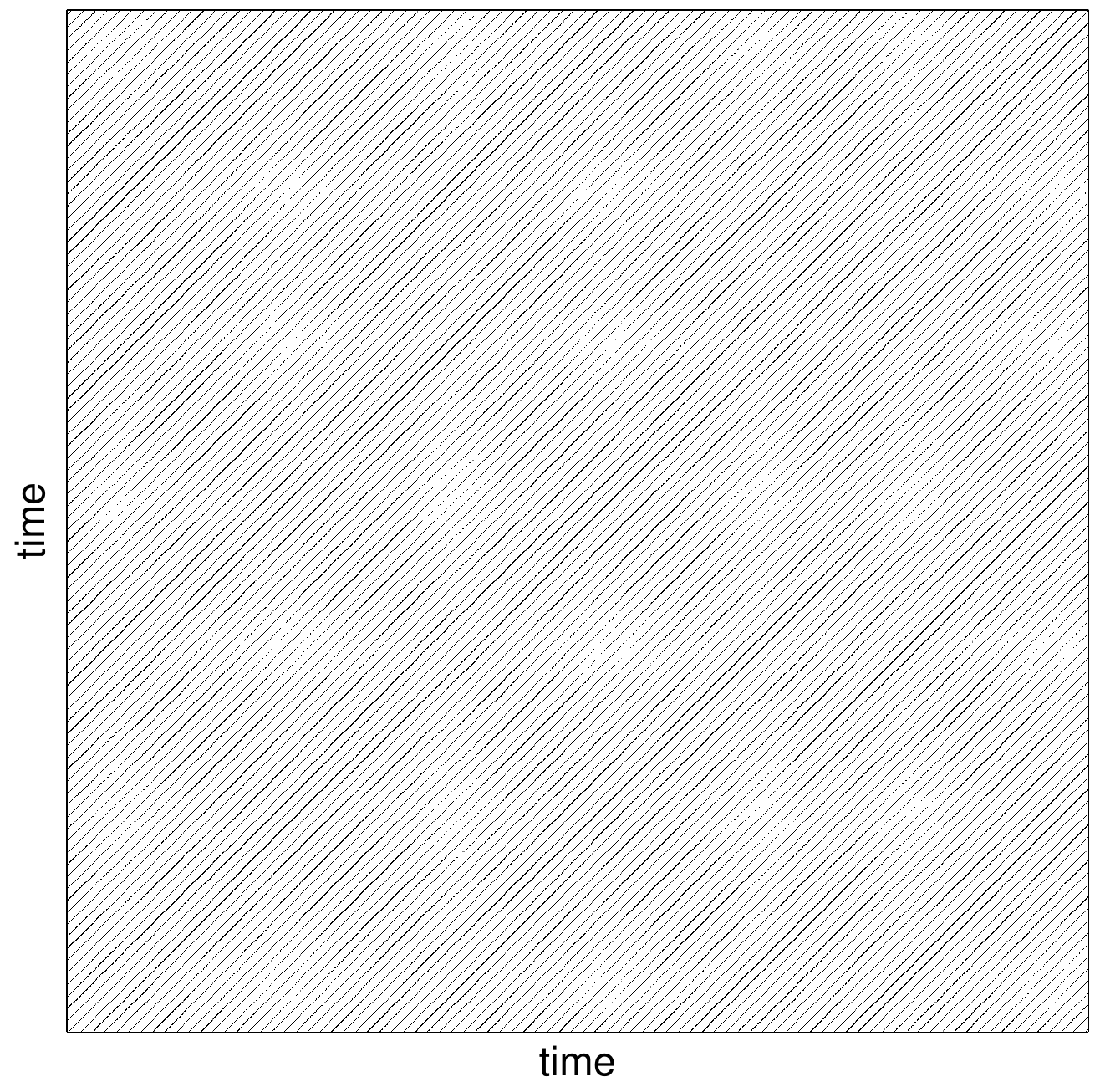}
    \includegraphics[width=0.3\textwidth]{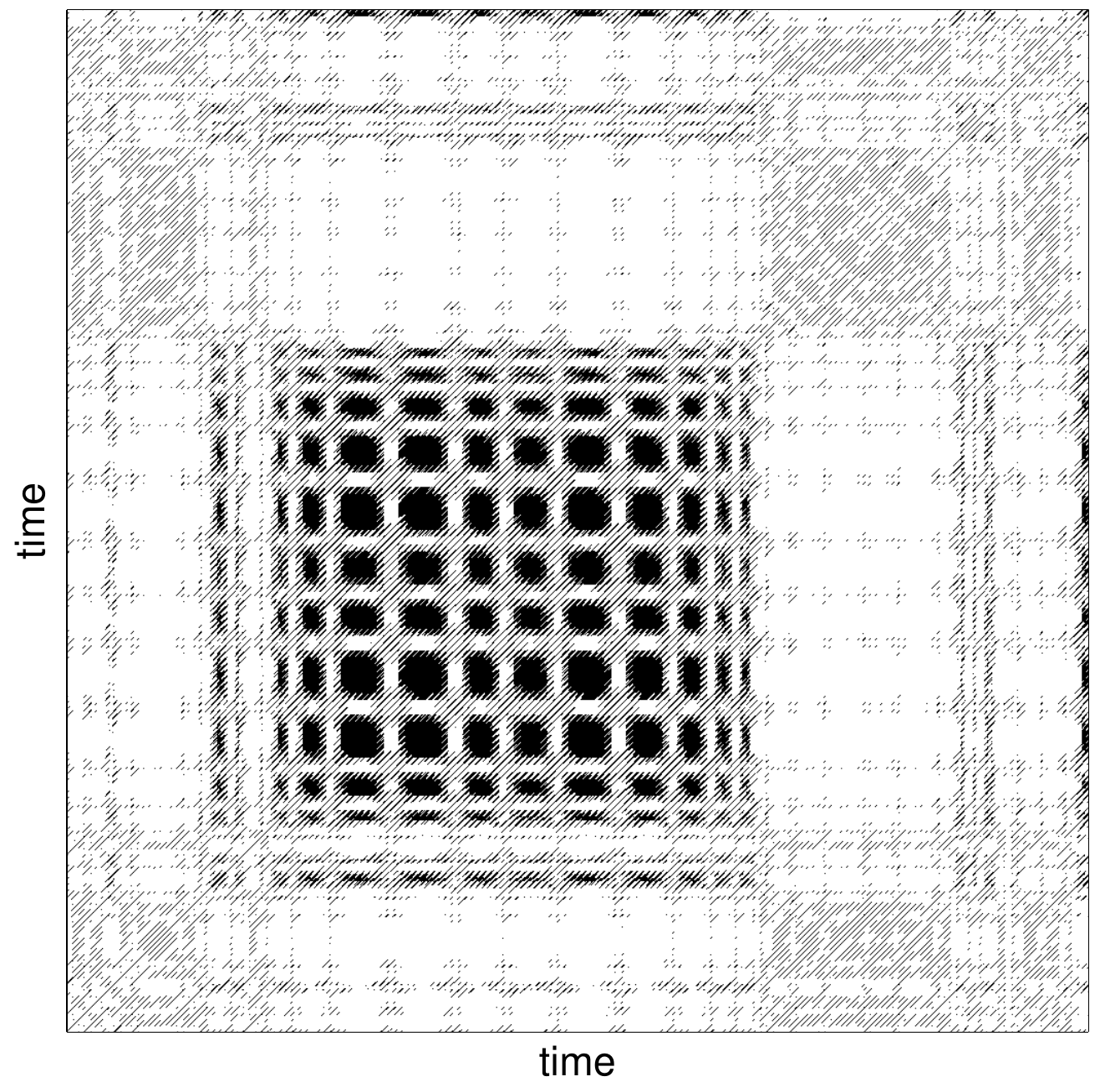}}
    \fi
    \caption[Recurrence plots of MPD orbits]{
    Recurrence plots of three MPD orbits. Left: regular with $S=0$ (geodesic), $E=0.976\mu$, $J_z=3.8\mu M$ and initial $r=4.5M$ ($\mathrm{RR}=0.08$, $d=2$ and $\mathcal{T}=6\Delta t$). Center: weakly chaotic at $S=10^{-4}\:\mu M$ with $E=0.976037\mu$, $J_z=3.8\mu M$ and initial $r=4.25216M$ ($\mathrm{RR}=0.08$, $d=20$ and $\mathcal{T}=4\Delta t$). Right: Strongly chaotic at $S=1.4\mu M$ with $E=0.92292941\mu$, $J_z=4.0\mu M$ and initial $r=4.5M$ ($\mathrm{RR}=0.15$, $d=7$ and $\mathcal{T}=\Delta t$). }
    
    \label{fig:mpd_rp1}
\end{figure*}

\begin{figure*}
    \centering
    \ifimages
        {\includegraphics[width=0.3\textwidth]{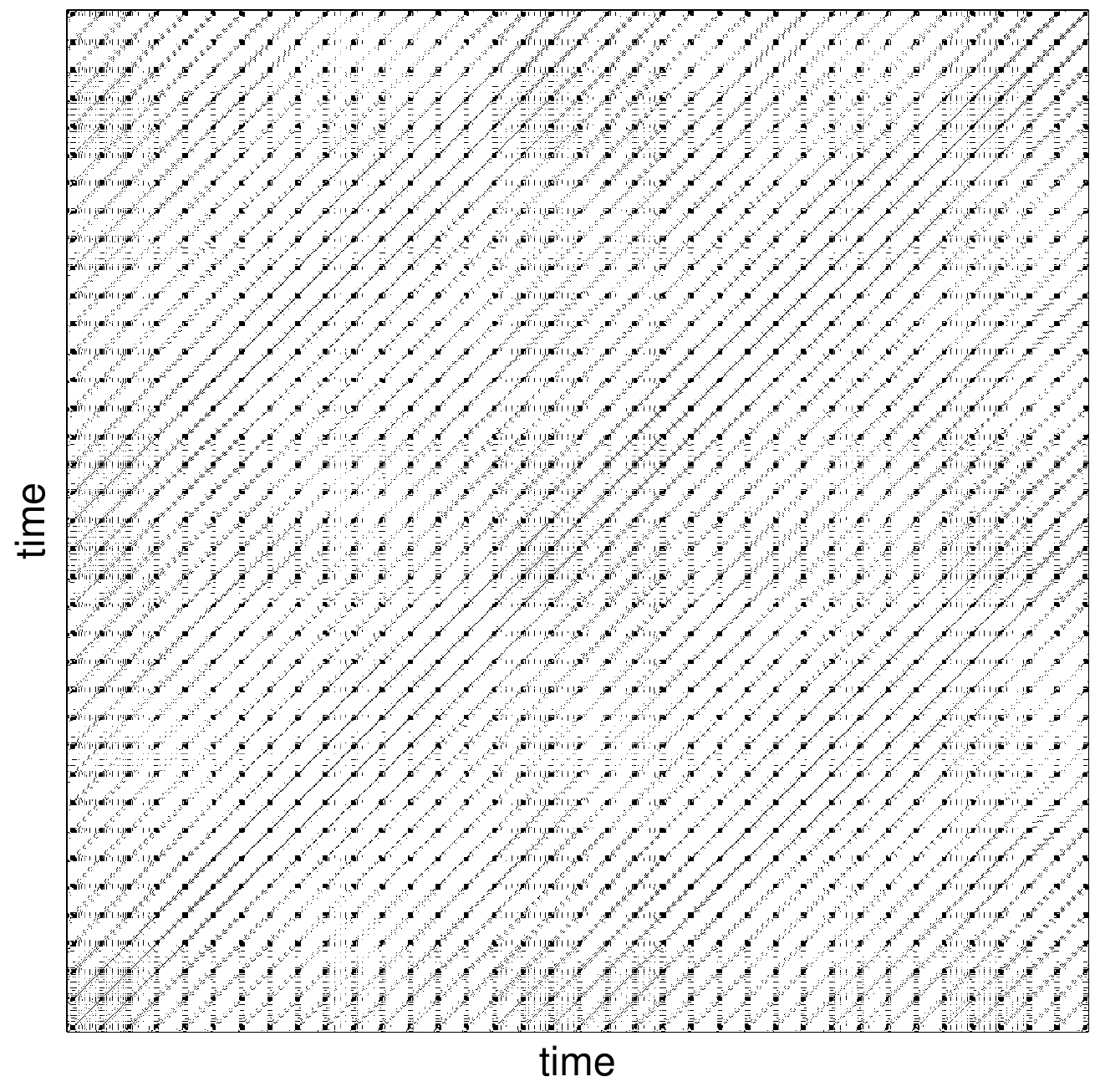}
        \includegraphics[width=0.3\textwidth]{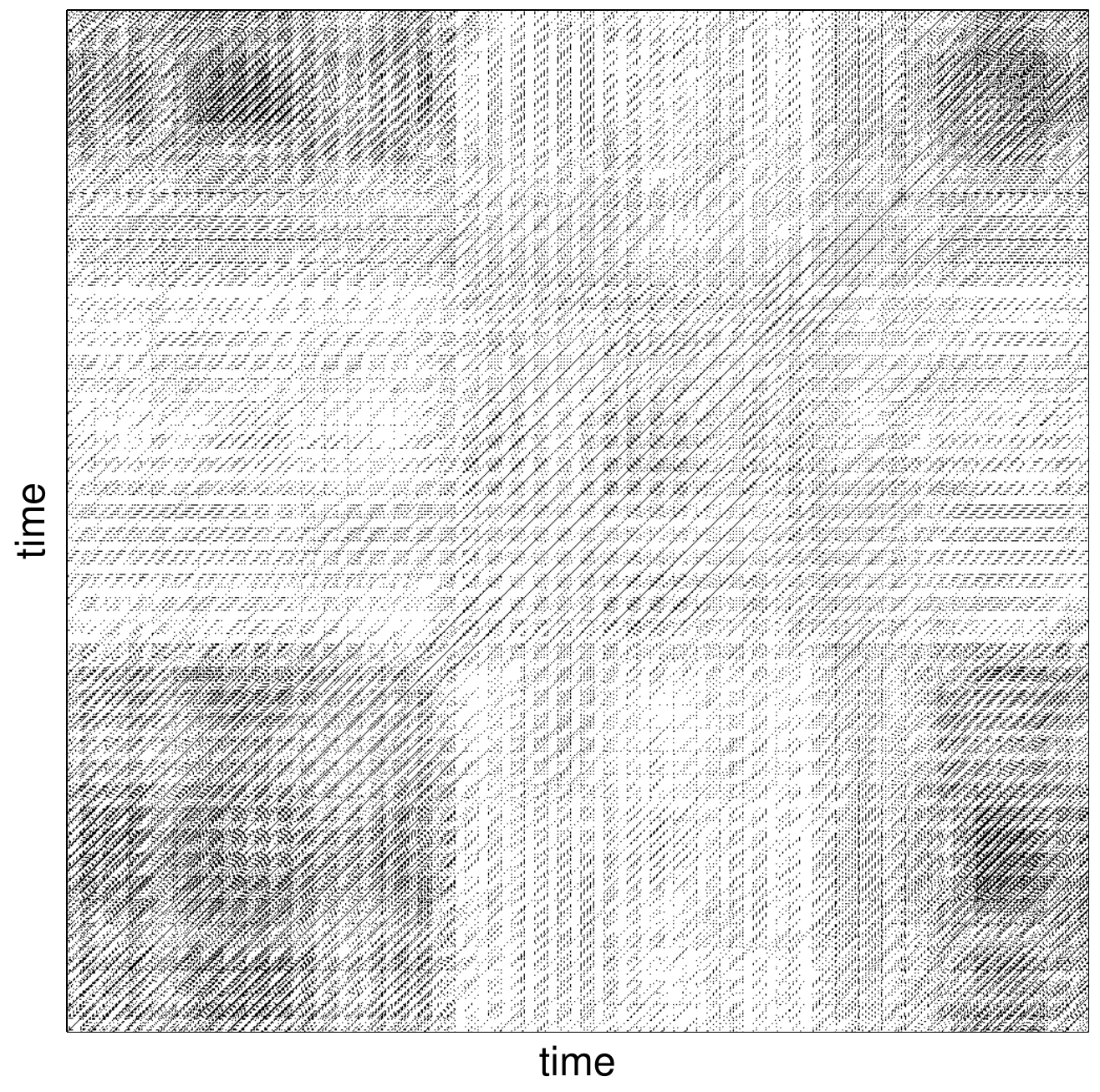}
        \includegraphics[width=0.3\textwidth]{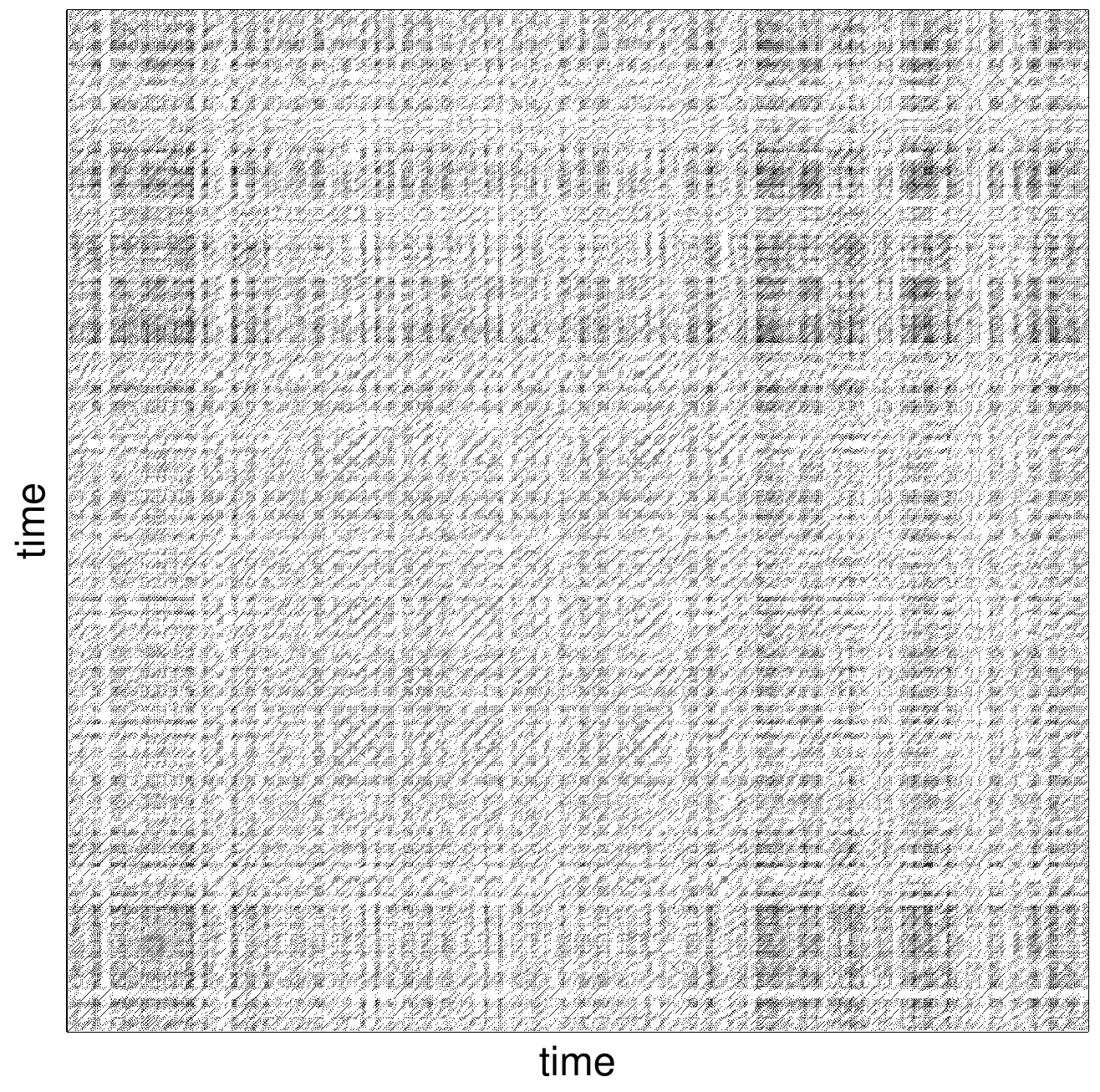}}
    \fi
    \caption[Recurrence plots of gravitational waveforms]{Recurrence plots of gravitational waveforms corresponding to the orbits of Fig.~\ref{fig:mpd_rp1}. Left panel shows the RP of the regular waveform ($\mathrm{RR}=0.08$, $d=4$ and $\mathcal{T}=7\Delta t$), middle panel represents the weakly chaotic case ($\mathrm{RR}=0.15$, $d=10$ and $\mathcal{T}=3\Delta t$) and RP of the waveform resulting from the strongly chaotic trajectory is presented in the right panel ($\mathrm{RR}=0.15$, $d=5$ and $\mathcal{T}=2\Delta t$).}
    \label{fig:mpd_wrp1}
\end{figure*}

\subsection{Recurrence plots}
Recurrence analysis is a method which allows us to analyze dynamics of regular and chaotic systems. It can be used to estimate dynamical invariants, such as the second-order R\'{e}nyi entropy and correlation dimension, to uncover features such as unstable periodic orbits and sticky orbits, or simply to distinguish linear and non-linear behavior. For more detailed information on recurrence analysis, including different metrics to quantify the recurrence plots for a more substantial analysis of time series, see \cite{Marwan07}.

All we require is a time series in the full phase space $\vec{x}_i$. We define the recurrence matrix as
\begin{equation}
    \mathbf{R}_{ij} = \left\{ \begin{array}{lr}
    \theta\left(\varepsilon - \left|\left|\vec{x}_i-\vec{x}_j\right|\right|\right) & i \neq j \\
    0 & i = j
    \end{array} \right. \:,
\end{equation}
where $\theta$ is the Heaviside step function and $\varepsilon$ is a free parameter called the recurrence threshold. Intuitively, this simply means: $\mathbf{R}_{ij} = 1 \Leftrightarrow$ there is a recurrence at times $i$ and $j \Leftrightarrow \vec{x}_i$ and $\vec{x}_j$ are closer than a given (small) threshold. The recurrence matrix is by definition symmetric. The main diagonal $i=j$ is excluded for technical reasons \cite{Marwan07}. There is also an implicit ambiguity in the choice of metric. We define the Recurrence Rate (RR) as the density of ones in the recurrence matrix, for a time series of length $l$:
\begin{equation}
    \mathrm{RR} = \frac{1}{l^2}\sum_{i,j=1}^{N,N}\mathbf{R}_{ij} \:.
\end{equation}
Often, a specific value of RR is chosen and the threshold $\varepsilon$ is then tuned to fit the given RR.

We can visualize this matrix in the graph with the two axes $i$ and $j$ corresponding to times $t_i$ and $t_j$, and representing ones in the recurrence matrix by dots in the graph. This is called a Recurrence Plot (RP). 

RP can be inspected visually to see some of the basic dynamical features, namely to distinguish linear and non-linear behavior. When a time series is quasiperiodic, there is a delay after which one finds recurrences for a sufficiently high recurrence threshold; i.e. $\forall i\: \mathbf{R}_{i, i+\Delta i}=1$. This forms lines parallel to the main diagonal offset by $\Delta i$ either in the horizontal or in the vertical direction.

In contrast, for a chaotic orbit, this regular structure only exhibits itself due to stickiness, a phenomenon where a chaotic orbit approaches an island of stability and mimics its regular behavior for a long period of time in interval $\mathcal{I}$. The part of RP in the region $\mathcal{I}\times \mathcal{I}$ is formed by diagonal-parallel lines and forms a regular-looking square lying on the main diagonal. Furthermore, if the orbit is in a sticky regime in the intervals $\mathcal{I}_1$ and $\mathcal{I}_2$, then the region $\left(\mathcal{I}_1\times \mathcal{I}_2\right) \bigcup \left(\mathcal{I}_2\times \mathcal{I}_1\right)$, which shows correlations between the intervals $\mathcal{I}_1$ and $\mathcal{I}_2$, can either show a similar structure again if the orbit moves near the same island during both intervals or be empty if the two islands are different. This forms the typical square-like structure of a recurrence plot.

\subsection{Time delay method}\label{sec:embedding}

The method of recurrence analysis is powerful, but it assumes that a trajectory in the phase space of dynamical system is available for inspection. However, it may happen that we only have access to limited data (e.g., a simple scalar time series representing a complex multi-dimensional system) and want to apply recurrence analysis; it is then necessary to use a phase space reconstruction technique. Here we describe the time delay embedding method based on Takens's theorem \citep{Takens81}.

Let us have a time series $\vec{x}_i$ of dimension $n_1$ and length $l_1 \geq i \in \N$ in space $\mathcal{X}$. Let us also choose two free parameters: time delay $\mathcal{T} \in \N$ and embedding dimension $d \in \N$. We then call $\mathcal{X}^d$ the reconstructed phase space and define the reconstructed time series as
\begin{equation}
    \vec{y}_i = \left(\vec{x}_i, \vec{x}_{i+\mathcal{T}}, ..., \vec{x}_{i+\mathcal{T}\left(d-1\right)}\right) \:,
\end{equation}
which has dimension $n_1\cdot d$ and is reduced in length to $l_1 - \mathcal{T}\left(d-1\right) = l_2 \geq i\in\N$. The trivial case without reconstruction is $d=1$.

This method has been shown by \citet{Takens81} to provide a diffeomorphism between the original and reconstructed phase space under some conditions. This means that even using the data from a single suitable (i.e. not constant along the system's evolution) scalar function on the phase space we can still study the system's dynamics - in fact, this removes some of the ambiguity of the metric choice.

It is also necessary to make a proper choice of parameters $\mathcal{T}$ and $d$. Time delay is usually chosen as the first minimum of either the mutual information or the autocorrelation function. Embedding dimension is determined by the False Nearest Neighbor (FNN) algorithm \citep{kennel92}. The phase space is reconstructed using a given time delay $\mathcal{T}$ and embedding dimension $d$ and for every point, the nearest point is found. It is considered a false nearest neighbor if their distance increases by a factor of at least $f$ (we typically use $f = 5 - 10$) when the embedding dimension is changed to $d+1$. A suitable $d$ is such that the fraction of FNNs is considered small enough (in ideal case it drops to zero).

\subsection{Recurrence plots of MPD orbits and corresponding gravitational waveforms}

First we have applied recurrence analysis to MPD orbits. In particular, we have sampled the radial coordinate $r$ along the orbit with constant time-step $\Delta t$ to obtain 1300 points and used this time series to construct the recurrence plots. Recurrence threshold $\varepsilon$ is determined by fixing the recurrence rate ($\mathrm{RR}$). For a geodesic case we set $\mathrm{RR}=0.08$, while for a chaotic orbit higher value $\mathrm{RR}=0.15$ appears appropriate. Embedding method is employed to reconstruct the trajectory in the phase space. Embedding dimension $d$ is determined for each time-series by the FNN algorithm as described above and embedding time delay $\mathcal{T}$ is found as a first minimum of time delayed mutual information.

We use the \texttt{CRP Toolbox} \cite{Marwan07} installed on \texttt{Matlab R2014a} to construct the RPs. Embedding parameters are estimated using the functions \texttt{false\_nearest} and \texttt{mutual} from  the \texttt{Toolbox of RP and RQA} \cite{chen12}. 

Computed recurrence plots are shown in Fig.~\ref{fig:mpd_rp1}. Note that their visual inspection reveals regular and chaotic behavior. In particular, for a geodesic case (left panel of Fig.~\ref{fig:mpd_rp1}) we observe simple diagonal pattern characteristic for regular dynamics. However, this simple structure of RP becomes distorted as the introduction of the small spin induces weak chaos (middle panel). With high spin value, we obtain typical RP of strongly chaotic system (right panel).

In the following, we apply the same procedure of recurrence analysis to the simulated gravitational waveforms corresponding to the MPD orbits. The waveforms were obtained numerically by the method described in Sec. \ref{sec:waves}. In particular, we use the time series $h_{+\:m\leq 3}$ as described in Subsec.~\ref{ssec:waveforms}. Analogically to the previous case, we use delay embedding method to reconstruct the phase space trajectory and the recurrence rate is fixed to determine the threshold $\varepsilon$.

The resulting recurrence plots of the waveforms are shown in Fig.~\ref{fig:mpd_wrp1}. The plots clearly confirm the onset of chaos and match RPs of the corresponding trajectories to a certain degree. In particular, we observe that chaotic behavior may be identified even for a weakly chaotic case with realistic value of spin (middle panel of Fig.~\ref{fig:mpd_wrp1}). Thus, we can conclude that chaos remains encoded in the gravitational waveforms radiated from EMRI and that it may be detected using the technique of recurrence analysis.\footnote{In cases of weakly chaotic layers, which usually are the case for low spin values, it is naturally harder to detect chaos and even low noise can be an obstacle, see also \cite{LukesGerakopoulos17b}}

\section{Conclusion}\label{sec:concl}

In the first part of this work, we have demonstrated that chaos manifests itself even for spin values relevant to real astrophysical events using numerically evolved orbits of spinning particles with the Tulczyjew-Dixon spin supplementary condition. We have designed and presented the fiducial action-angle variables and a method of studying any resonance for any kind of perturbation of the conventional geodesic model in the Schwarzschild spacetime.

It is known that the inclusion of spin in the equations of motion causes the system to lose integrability \cite{Hartl02}; however, due to the existence of approximate constants of motion \cite{Ruediger81}, it is not yet clear whether this means that there are no prolonged resonances at linear order. This work shows that the $1/2$ and $2/3$ resonances are related to the second-order in spin terms and do not manifest themselves at linear order in spin in the Schwarzschild spacetime. Thus, we have provided evidence that terms linear in spin do not cause the emergence of chaos, which in turn has little effect on the overall dynamics of an EMRI with a non-spinning supermassive black hole.

This result supports the expectation that spin-induced chaos and resonances will not play a significant role in EMRIs. Our work provides numerical evidence for the case of two most significant $r-\theta$ resonances in the Schwarzschild spacetime. It is of great interest to extend this analysis to the Kerr spacetime to get a more complete picture of the dynamics of an EMRI in the more general case with a spinning primary. It would also be beneficial to show that our conclusion is independent of the spin supplementary condition used.

In the second part of the work, we have used the Teukode, a time-domain Teukolsky equation solver, to compute gravitational waveforms generated by the motion of a spinning particle in the Schwarzschild background. Using recurrence plots, we have established a close link between dynamical features of the particle's motion and the corresponding gravitational waveforms with high enough spins. For low spins, however, we have to elaborate to confirm this connection. For example, one has to eliminate the numerical noise from the Teukode simulation. It would be of interest to repeat the simulations with a finer grid so as to see whether the reduced noise would contribute to a closer link between the orbit's and waveform's recurrence plots. Another problem that could be addressed is the nature of recurrence analysis, which is by design suited to systems with a finite number of degrees of freedom. Finally, a large challenge is the fact that the real signal will not only be heavily distorted by background and detector noise \cite{LukesGerakopoulos17b}, but it will also be overlapping with signals from other sources.

Even though it is yet not clear whether recurrence analysis is able to detect spin-induced chaos in EMRI, there is an ongoing effort to establish it as a tool able to detect chaos from other sources. For example, in the case that chaos arise due to a broken spacetime background symmetry, which assumes that the Kerr hypothesis is incorrect (see, e.g., \cite{LukesGerakopoulos17b}) and prolonged resonances can occur due to the gravitational perturbation by the surrounding astrophysical environment \cite{Bonga19}.

\begin{acknowledgments}
The authors have been supported by the Grant No. GA\v{C}R-17-06962Y of the Czech Science Foundation, would like to acknowledge networking support by the COST Action CA16104 and Inter-Excellence COST project LTC 18058 of the Czech Ministry of Education, Youth and Sports. G.L.-G. would also like to express gratitude for the hospitality of the Theoretical Physics Institute at the University of Jena. Finally, we would like to thank Sebastiano Bernuzzi, Enno Harms and Sarp Akcay for useful discussions and comments. 
\end{acknowledgments}

\appendix

\section{Analytic properties of geodesics}\label{sec:geo}

When considering geodesic motion in the Schwarzschild spacetime, the transformation between the coordinates $\left(x^\mu,P_\nu\right)$ and the AA variables can be explicitly written using integrals, for full derivation see \cite{Fujita09}. We only give the $r$-action and angle. They are given using the polynomial $R$ and the integral $\mathcal{R}$, which are defined as
\begin{subequations}
\begin{align}
    R\left(r;E,J_z,\mathcal{C}\right) &= \frac{E^2 r^4}{\mu^2} - r^2 f\left(r^2+J_z^2+\mathcal{C}\right) \:, \\
    \mathcal{R}\left(r\right) &\defeq \int_{r_2}^{r}R^{-1/2}\left(r'\right)\d r' \:,
\end{align}
\end{subequations}
as
\begin{subequations}
\begin{align}
    & I_r^{geo}\left(r;E,J_z,\mathcal{C}\right) = 2\int_{r_2}^{r_1}R^{1/2}\left(r'\right)\d r' \:, \label{eq:action_geo} \\
    \begin{split}
    & \theta^r_{geo}\left(r;E,J_z,\mathcal{C}\right) = \\
    &\left\{ \begin{array}{lr}
        \pi\mathcal{R}\left(r\right)/\mathcal{R}\left(r_1\right)  & P_r\geq 0\\[0.5em]
        -\pi\mathcal{R}\left(r\right)/\mathcal{R}\left(r_1\right)   &  P_r < 0
    \end{array}\right. \,, \label{eq:angle_geo}
    \end{split}
\end{align}
\end{subequations}
where $r_1$ and $r_2$ are turning points; they are the highest and next-to-highest root of $R\left(r\right)$, respectively. The integrals are the energy $E$, the azimuthal component of the orbital angular momentum $J_z$ and the Carter constant $\mathcal{C}$, which in the Schwarzschild case reduces to $\mathcal{C} = J_x^2 + J_y^2$, and their presence in $R$ is implied. This way, the angle coordinate is defined with the convention $\theta^r_{geo} \in \left[-\pi, \pi\right]$.

The integrals in $\theta^r_{geo}$ can also be expressed using special functions, namely the incomplete elliptic integral of the first kind
\begin{align}\label{eq:ellipkinc}
    \begin{split}
    F\left(\varphi,k\right) &= \int_0^\varphi \frac{\d\vartheta}{\sqrt{1-k^2\sin^2\!\vartheta}} = \\
    &= \int_0^{\sin\varphi}\frac{\d y}{\sqrt{\left(1-y^2\right)\left(1-k^2y^2\right)}}
    \end{split}
\end{align}
and the complete integral of the first kind $K\left(k\right) = F\left(\pi/2,k\right)$.

It is first necessary to find all the roots of $R\left(r\right)$ and rewrite as
\begin{align}
    \begin{split}
    R\left(r\right) = \left(1-\frac{E^2}{\mu^2}\right)&\left(r_1-r\right)\left(r-r_2\right)\left(r-r_3\right)r, \\
    &0 \leq r_3 \leq r_2 \leq r_1 \:.
    \end{split}
\end{align}
Then we can express
\begin{align}
    \begin{split}
    \int_{r_2}^{r} R^{-1/2}\left(r'\right)\d r' =& \\ \frac{2}{\left(1-E^2/\mu^2\right)\left(r_1-r_3\right)\left(r_2-r_4\right)} & F\left(\arcsin y_r,k_r\right) \:,
    \end{split}
\end{align}
where
\begin{subequations}
\begin{align}
    y_r &= \sqrt{\frac{r_1-r_3}{r_1-r_2}\frac{r-r_2}{r-r_3}} \:, \\
    k_r &= \sqrt{\frac{r_1-r_2}{r_1-r_3}\frac{r_3-r_4}{r_2-r_4}} \:.
\end{align}
\end{subequations}
Thus, the angle variable can be expressed as
\begin{equation}\label{eq:angle_r}
    \theta^r_{geo} = \left\{ \begin{array}{lr}
        \pi F(\arcsin y_{r},k_{r})/K(k_{r}) & P_r\geq 0\\[0.5em]
        -\pi F(\arcsin y_{r},k_{r})/K(k_{r}) & P_r < 0
    \end{array}\right. \:.
\end{equation}
The action variable, however, employs an integral for which there does not seem to be a closed form expression even using special functions.

To perform the conversion, we make use of Python 3 and the \verb|numpy| and \verb|scipy| libraries. More specifically, the function \verb|scipy.integrate.quad| was used to compute the integral in Eq. \eqref{eq:action_geo} and \verb|scipy.special.ellipkinc| was used to evaluate the elliptic integral (Eq. \eqref{eq:ellipkinc}) in order to compute the angle variable in Eq. \eqref{eq:angle_geo}.

\section{Numerical solutions of the MPD equations}\label{sec:mpd_code}

For numerical integration of orbits of spinning particles, a Fortran code was written. The input are values for initial $r$, $P_r$ and integrals $E$, $J_z$ and $S$; the code places a spinning particle with these parameters in the equatorial plane and calculates all the coordinate components of the four-momentum and the spin tensor so as to satisfy the input values, $J_x = J_y = 0$, and the TD SSC. In this part, real variables are represented as quadruple precision floating point variables due to the ill-conditioned calculation of expressions such as $E+P_t$ and $J_z - P_\phi$.

Then it evolves the MPD equations using the Gauss collocation method of different orders using fixed-point iteration. Here, real variables are represented as double precision floating point. In this work, 4th order was used for all computations. The rotation number is computed using Eq. \eqref{eq:angular_moment} for each initial condition.

\begin{figure}
    \centering
    \ifimages
     \includegraphics[width=0.45\textwidth]{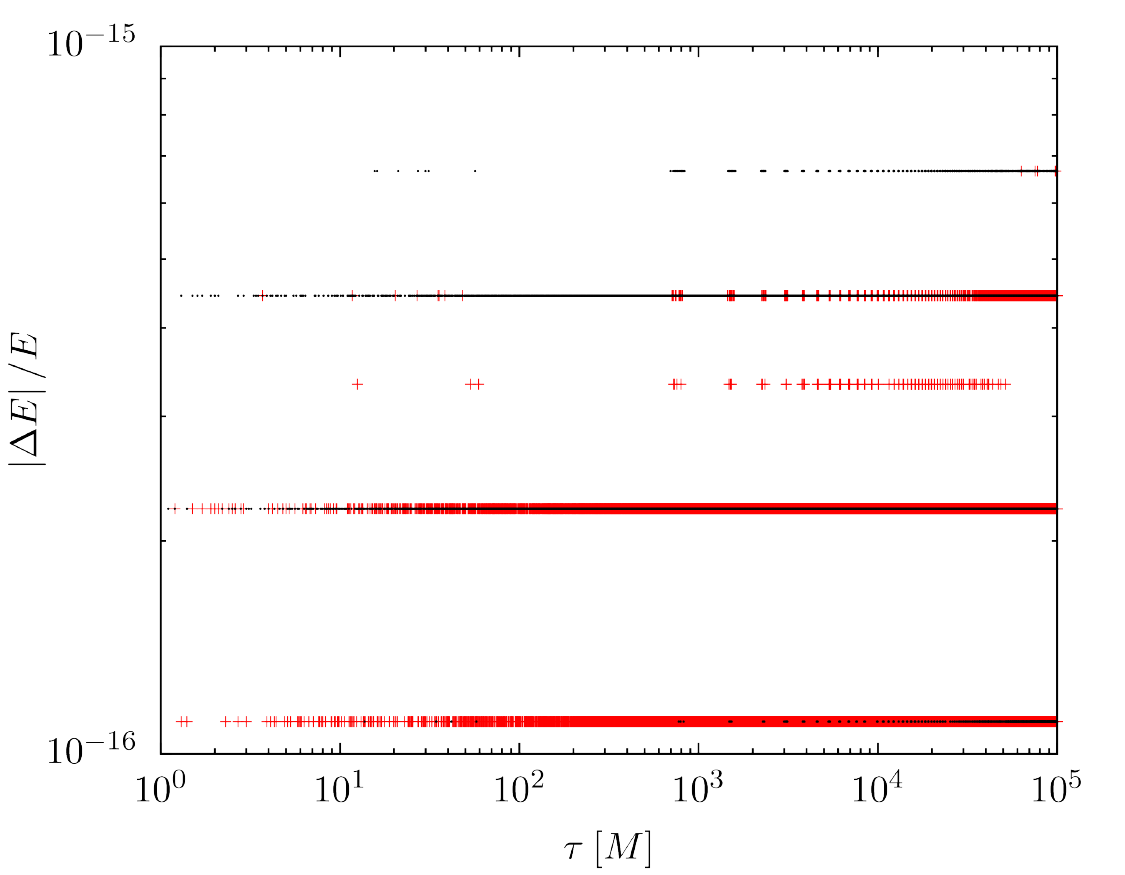}
    \fi
    \caption[Integration error of MPD equations]{Evolution of the relative error $\Delta E = \abs{E\left(\tau\right)-E\left(0\right)}/E\left(0\right)$ for the trajectories of Fig.~\ref{fig:mpd_wav2}. Black points correspond to the regular trajectory, red crosses correspond to the chaotic trajectory.}
    \label{fig:err1}
\end{figure}

The code deals with the system as originally written in Eq. \eqref{eq:mpd}, i.e. $x^\mu$, $P_\mu$ and $S^{\mu\nu}$; no reduction is applied for the integration. Thus, the integrals $E$, $J_i$, $S^2$, $\mu^2$ and $P_\mu S^{\mu\nu}$ are used to track the integration error. In Fig. \ref{fig:err1}, evolution of the relative error of energy is shown to be very low - on the order of $10^{-14}$. 

Slightly different versions of the code were also used to locate the unstable periodic orbit and a given resonance by evolving the equations of motion and using a bisection method to solve an equation employing a function of the initial $r$ along the $P_r=0$ line. In the first case, the corresponding equation is $P_r = 0$ at the next intersection with the section; in the second, it is $\nu_\vartheta=\omega$, where $\omega$ is the desired rotation number. To make sure that the accuracy grows as the code converges, the number of iterations of the return mapping grows as well: if the two current estimates are $\omega_1 > \omega_2$, the next orbit is integrated long enough as to get $2/\mathrm{min}\left(\omega_1 - \omega_{target}, \omega_{target} - \omega_2 \right)$ intersections.

To determine the width of a resonance $p/q$, Python 3 with the \verb|numpy| and \verb|scipy| libraries was used to fit Eq.~\eqref{eq:pert_to_wid}
to Poincar\'{e} section points (see Eq. \eqref{eq:ang_to_wid}). Only points with $\abs{\theta^r} < 0.01 \cdot 2\pi/(ns)$ were taken into account and as components of the vector of residuals we use
\begin{equation}\label{eq:fit_width}
-I_r + I_{r}^0 + \mathrm{sign}\left(I_r - I_r^0\right)\frac{\mathrm{width}\cdot ns}{4}\theta^r \:.
\end{equation}
This expression together with the function \verb|scipy.optimize.least_squares| was used to estimate the values of $I_r^0$ and the $\mathrm{width}$.

\begin{figure}
    \centering
    \includegraphics[width=0.45\textwidth]{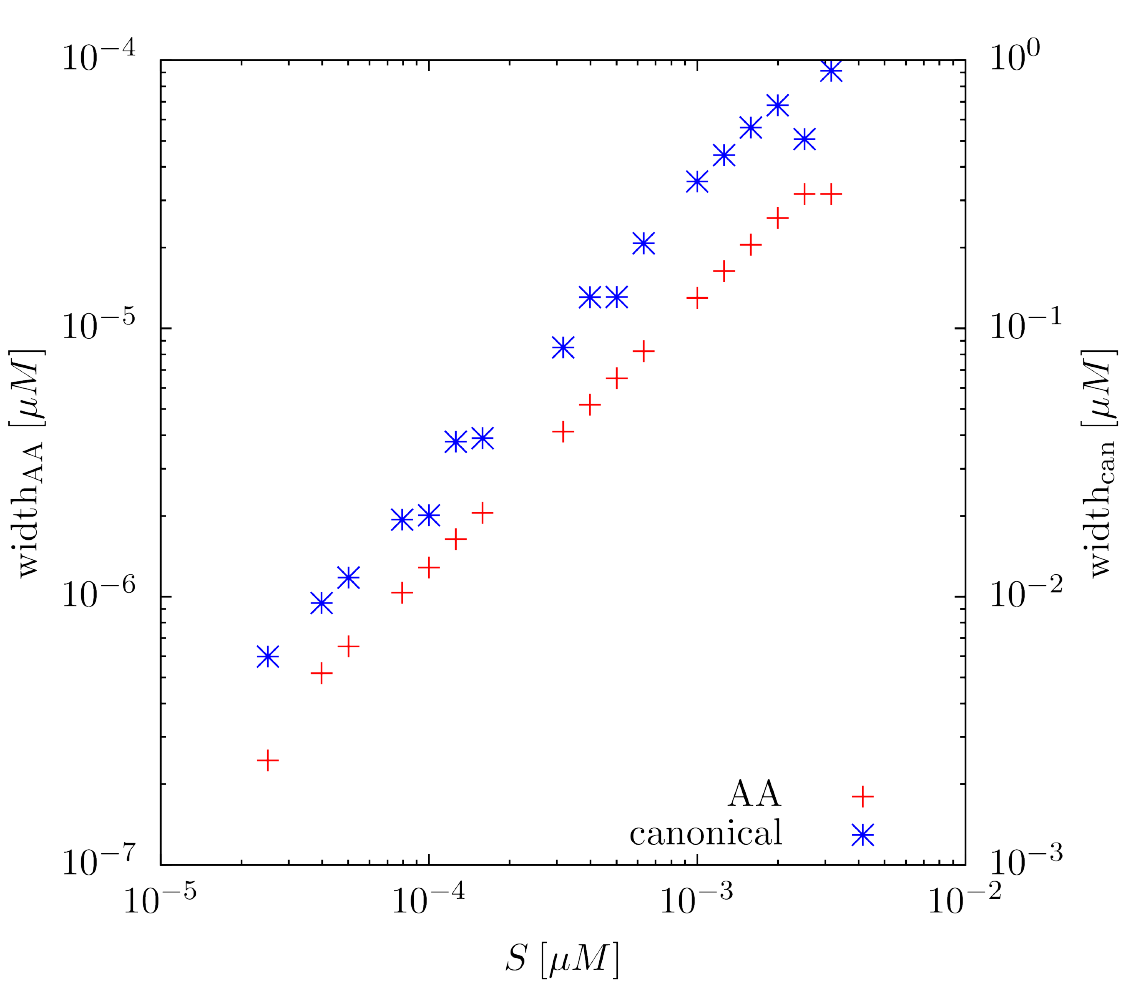}
    \caption[Comparison of AA and canonical coordinates]{Comparison of resonance growth in action-angle variables (left vertical axis) and in canonical coordinates (right vertical axis).}
    \label{fig:mpd_gro2}
\end{figure}

To demonstrate the advantage of action-angle variables, the same process was applied to estimate the width of the resonance in canonical $r, P_r$ coordinates for the $2/3$ resonance. The function to fit is again in Eq.~\eqref{eq:fit_width}, with $r$ taking the role of $I_r$ and $P_r$ of $\theta_r$. Fig.~\ref{fig:mpd_gro2} compares the performance of the two coordinate sets and demonstrates that while the trend is undeniable in both, it is clearer in the AA variables. 

\section{Teukode}\label{sec:teukode}

\subsection{Horizon-penetrating, hyperboloidal coordinates}\label{sec:hh}

The Teukode has been written in Jena as part of an MSc. thesis and the following dissertation \cite{Harms16}; for a more concise summary, see \cite{Harms14}. It is a time-domain solver for the master equation \eqref{eq:master} with a point particle source. 
Different coordinate systems are used in the Teukode in order for the equation to be regular at the horizon and to smoothly reach the future null infinity $\mathcal{J}^+$. The system of choice is the horizon-penetrating, hyperboloidal (HH) coordinate system, see \cite{Yang13}. Here, its construction is described in the simpler Schwarzschild case.

The first step is to define
\begin{equation}
    \tilde{t} = t + 2M\log\abs{r-2M} \:.
\end{equation}
Then, one uses the technique of hyperboloidal compactification. The new variables $\HHT$, $\rho$ are defined by
\begin{subequations}
\begin{align}
    \rho\left(r\right) &\vcentcolon\quad r = \frac{\rho\left(r\right)}{\Omega\left(\rho\left(r\right)\right)} \:, \\
    \HHT\left(\tilde{t},\rho\right) &\defeq t - h\left(\rho\right) \:,
\end{align}
\end{subequations}
where the $h$ is called the height function and $\Omega$ the conformal factor. The choice made for them here is
\begin{subequations}
\begin{align}
    \Omega\left(\rho\right) &\defeq 1 - \frac{\rho}{\Sigma} \:, \\
    h\left(\rho\right) &\defeq \frac{\rho}{\Omega} - \rho - 4M\log\Omega \:,
\end{align}
\end{subequations}
where $\Sigma$ is a free parameter and the location of $\mathcal{J}^+$ then corresponds to $\rho = \Sigma$. The position of the horizon is then
\begin{equation}
    \rho_+ = \frac{2M\Sigma}{2M+\Sigma} \:.
\end{equation}

\subsection{Numerical solutions of the Teukolsky equation}

The Teukode uses the HH coordinates to smoothly cover the whole region of interest from the horizon all the way to the null infinity $\mathcal{J}^+$. The equation is separated into m-modes by taking the Fourier transform in the $\phi$-direction
\begin{equation}
    \psi_4\left(\HHT, \rho, \theta, \phi\right) = \sum_{\mathrm{m}=-\infty}^{\infty} \Psi_{\mathrm{m}} e^{i\mathrm{m}\phi} \:,
\end{equation}
the master equation \eqref{eq:master} is then of the form
\begin{align}\label{eq:master_sep}
    \begin{split}
    &C_{\HHT\HHT}\partial_{\HHT\HHT}\Psi_\mathrm{m} + C_{\HHT\rho}\partial_{\HHT\rho}\Psi_\mathrm{m} + C_{\rho\rho}\partial_{\rho\rho}\Psi_\mathrm{m} \\ &+ C_{\theta\theta}\partial_{\theta\theta}\Psi_\mathrm{m} + C_\HHT\partial_\HHT\Psi_\mathrm{m} + C_\theta\partial_\theta\Psi_\mathrm{m} \\ &+ C_\rho\partial_\rho\Psi_\mathrm{m} + C_0\Psi_\mathrm{m} = S_{-2} \:.
    \end{split}
\end{align}

The Teukode uniformly discretizes the interval $\left[\rho_+, \Sigma\right] \times \left[0, \pi\right]$ and implements different finite difference stencils up to 8th order to transform the reformulated 2+1 Teukolsky equation \eqref{eq:master_sep} into a set of ODE; we used 6th order finite differencing. This is then evolved using a standard 4th order Runge-Kutta method. The step size is determined using the CFL condition as $\Delta t = C_{CFL}\min\{h_\rho, h_\theta\}$, where the $h_\rho$ and $h_\theta$ are the spacing in the $\rho$ and $\theta$ direction, respectively, and $C_{CFL}\geq 1$. In this work, we use $C_{CFL} = 2$.

\begin{figure}
    \centering
    \ifimages
     \includegraphics[width=0.45\textwidth]{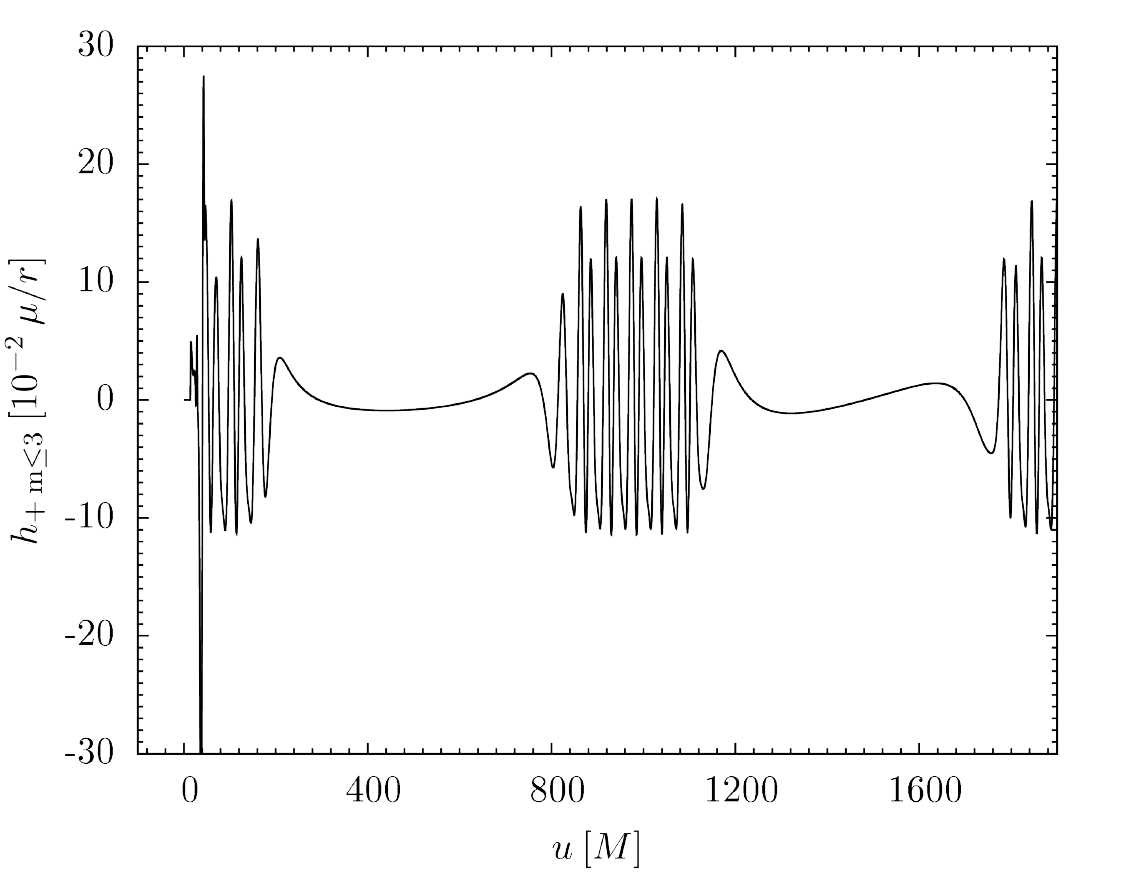}
    \fi
    \caption[Initial error of a Teukode simulation]{Output of the initial $1900M$ of the Teukode simulation corresponding to the chaotic orbit of Fig. \ref{fig:mpd_wav2}.}
    \label{fig:mpd_wavin1}
\end{figure}

As an initial value problem, solving the Teukolsky equation requires initial conditions as well. Interference of the radiated waves from the previous $\sim 2$ orbits is quite important here, so we have chosen the simplest possible way: to use $\psi_4 = 0\: \forall \rho, \theta$ and discard the first $\sim 200M$. The contrast of this beginning and later part is shown in Fig. \ref{fig:mpd_wavin1}.

The source term is a complicated linear combination of the stress-energy tensor components. For a spinning particle, the stress energy tensor is only non-zero in a single point of the grid, represented by a $\delta$ function and its derivatives up to 3rd order. This is modeled using a narrow Gaussian approximation. For a non-spinning particle source, the Teukode is also equipped with $n$-point delta approximations.

The Teukode also calculates energy and angular momentum fluxes.
In the eccentric orbits presented here for $S=10^{-4}\: \mu M$, the energy losses are $\dot{E} \doteq 10^{-3}\: \mu^2/M$ and angular momentum losses $\dot{J}_z \doteq 10^{-2}\: \mu^2$. This means that in our longest Teukode integration $t = 8.9\cdot 10^{4}\:M$, assuming $\mu/M \doteq 10^{-4}$, we get a total relative energy loss $\Delta E/E \doteq 1\%$ and total relative angular momentum loss $\Delta J_z/J_z \cdot 2.3\%$.

Also, the advanced time $v$ and retarded time $u$ are defined as
\begin{subequations}
\begin{align}
    u\left(t,r\right) &\defeq t - r_\ast \:, \\
    v\left(t,r\right) &\defeq t + r_\ast \:, \\
    r_\ast\left(r\right) &\defeq r + 2M\log\left(\frac{r}{2M} - 1\right) \:.
\end{align}
\end{subequations}
Their meaning is that for an outgoing radial null geodesic (i.e. $\dot{\theta} = \dot{\phi} = 0$, $\dot{r} > 0$, $\d s^2 = 0$) $u$ is a constant and only $v$ changes along the geodesic - this makes $u$ an ideal parameter for a waveform extracted at $\mathcal{J}^+$.

\begin{figure}
    \centering
    \ifimages
     \includegraphics[width=0.45\textwidth]{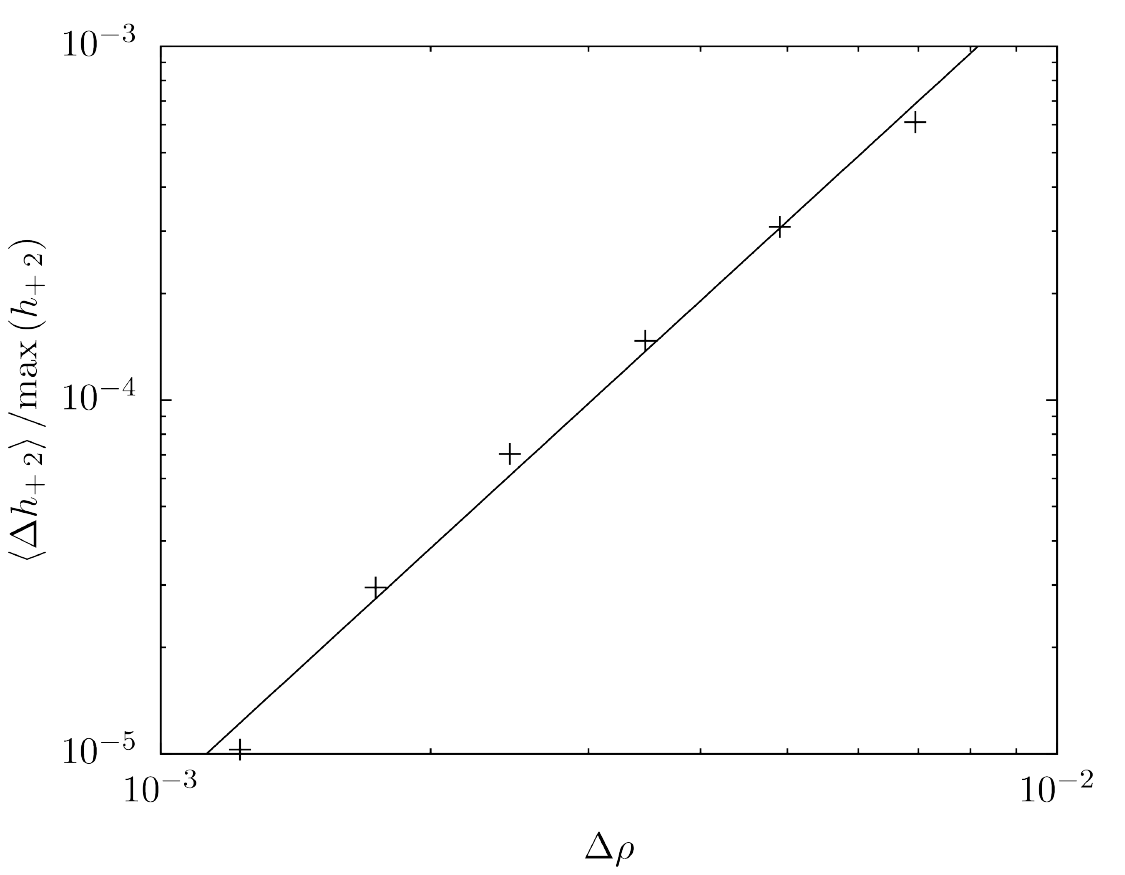}
    \fi
    \caption[Teukode convergence plot]{Convergence plot of the Teukode.}
    \label{fig:err2}
\end{figure}

To check the calculation errors, we have run simulations with different grids on the same chaotic orbit as shown in Fig. \ref{fig:mpd_wav2}. The waveform was calculated with grids $1201\times 61$, $1701\times 141$, $2401\times 201$, $3401\times 281$, $4801\times 401$, $6801\times 561$ and $9601\times 801$; of these, the finest $9601\times 801$ was taken as reference (closest to the exact solution, which we cannot get in any other way) and errors of the calculations with coarser grids were computed with respect to the reference grid. We calculated the averaged absolute error
\begin{equation}
    \left<\Delta h_{+\: 2}^{\mathrm{nx}} \right> = \frac{1}{50M}\int_{160M}^{210M} \abs{h_{+\: 2}^{\mathrm{nx}} - h_{+\: 2}^{9601}} \d u \:,
\end{equation}
where $h_{+\: 2}^{\mathrm{nx}}$ is the $+$ polarization of the $\mathrm{m}=2$ mode of the waveform calculated using $\mathrm{nx}$ points in the $\rho$ direction distributed uniformly from the horizon at $\rho = 5/3$ and the null infinity $\mathcal{J}^+$ at $\rho = 10$, extracted at $\mathcal{J}^+$. The convergence plot is shown in Fig. \ref{fig:err2} with the grid spacing on the horizontal axis and the vertical axis showing the relative error with respect to the maximum of the reference waveform in the given interval.

The line corresponds to a fit of the function
\begin{equation}
    \log \frac{\left<\Delta h_{+\: 2}\right>}{\mathrm{max}\left(h_{+\: 2}\right)} = A + q\cdot\log\left(\Delta\rho\right) \:,
\end{equation}
which was performed in logscale and returned the values
\begin{subequations}
\begin{alignat}{2}
    A &= 4.3  &&\pm 0.6 \:, \\
    q &= 2.32 &&\pm 0.09 \:.
\end{alignat}
\end{subequations}

Since the evolution method used is 4th order, one would expect $q=4$. However, the narrow Gaussian approximation for $\delta$ functions and their derivatives up to 3rd order is another source of error \cite{Harms16}.

The simulations in the text were carried out on the Virgo cluster at the Astronomical Institute of the Czech Academy of Sciences and split into multiple processes using MPI; typically, the $\rho$-direction was kept as a whole and the division was done into 8, 16 or 32 processes along the $\theta$-coordinate line. 

In the calculations carried out for this work, the HH coordinates were used with $\Sigma = 10$. The time step was determined using the Courant-Friedrichs-Lewy condition with $C_{CFL} = 2$. For more details on the coordinate system and numerical method, see Appendix~\ref{sec:teukode}.

\bibliographystyle{unsrt}
\bibliography{refs}

\end{document}